\documentclass[useAMS,usenatbib]{mnras}
\usepackage{epsfig}
\usepackage{color} 
\usepackage{natbib}
\usepackage{graphicx}
\usepackage{amsmath,amssymb}
\usepackage{amssymb}
\usepackage{float,lscape}
\usepackage{longtable}
\def\ew{$W_{r}$}

\def\hi{H~{\sc i}~} 
\def\nhi{$N${\sc (H~i)}} 
 
\def\mgii{Mg~{\sc ii}~} 
 
\def\siia{Si~{\sc ii}$\lambda$1526} 
\def\sii{Si~{\sc ii}}

\def\feii{Fe~{\sc ii}~}

\def\ovi{O~{\sc vi}~}

\def\civ{C~{\sc iv}~}

\def\zabs{$z_{\rm abs}$} 
\def\zem{$z_{\rm em}$~} 
 
\def\lya{Ly$\alpha$~} 
\def\llya{$L_{\rm Ly\alpha}$~} 
\def\lyb{Ly$\beta$~}

\def\kms{km s$^{-1}$} 
\def\aj{{AJ}}%
\def\araa{{ARA\&A}}%
\def\apj{{ApJ}}%
\def\apjl{{ApJ}}%
\def\apjs{{ApJS}}%
\def\aap{{A\&A}}%
\def\aapr{{A\&A~Rev.}}%
\def\jcap{{J. Cosmology Astropart. Phys.}}%
\def\mnras{{MNRAS}}%
\def\pasa{{PASA}}%
\def\pasp{{PASP}}%
\title[The \lya emission from DLAs]{ 
The \lya emission from high-$z$ galaxies hosting  strong Damped \lya systems}
\author[Joshi, R. et al.]{Ravi Joshi$^{1}$\thanks{E-mail: rjoshi@iucaa.in(RJ)}, Raghunathan Srianand $^{1}$, Pasquier Noterdaeme$^{2}$ and  Patrick  Petitjean$^{2}$  \\
$^{1}$Inter-University Centre for Astronomy and Astrophysics, Post Bag 4, Ganeshkhind, Pune 411007, India \\
$^{2}$UPMC-CNRS, UMR7095, Institut d'Astrophysique de Paris, F$-$75014 Paris, France \\
}

\begin{document}
\date{Accepted ---. Received ---; in original form ---}

\pagerange{\pageref{firstpage}--\pageref{lastpage}} \pubyear{2016}

\maketitle

\label{firstpage}
\begin{abstract}

We study the average \lya emission associated with high-$z$ strong
(log~\nhi~$\ge$~21) damped \lya systems (DLAs). We report \lya
luminosities (\llya) for the full as well as various sub-samples based
on \nhi, $z$, $(r-i)$ colours of QSOs and rest equivalent width of
\sii$\lambda$1526 line (i.e., $W_{1526}$). For the full sample, we
find \llya$<~10^{41}~(3\sigma)~\rm~erg~s^{-1}$ with a $2.8\sigma$
level detection of \lya emission in the red part of the DLA trough.
The \llya is found to be higher for systems with higher $W_{1526}$
with its peak, detected at $\geq~3~\sigma$, redshifted by about
300-400~\kms\ with respect to the systemic absorption redshift, as
seen in Lyman Break Galaxies (LBGs) and \lya emitters. A clear
signature of a double-hump \lya profile is seen when we consider
$W_{1526}~\ge~0.4$~\AA\ and $(r-i)~<~0.05$. Based on the known
correlation between metallicity and $W_{1526}$, we interpret our
results in terms of star formation rate (SFR) being higher in high
metallicity (mass) galaxies with high velocity fields that facilitates
easy \lya escape. The measured \lya surface brightness requires local
ionizing radiation that is 4 to 10 times stronger than the
metagalactic UV background at these redshifts. The relationship
between the SFR and surface mass density of atomic gas seen in DLAs is
similar to that of local dwarf and metal poor galaxies. We show that
the low luminosity galaxies will contribute appreciably to the stacked
spectrum if the size-luminosity relation seen for \hi at low-$z$ is
also present at high-$z$. Alternatively, large \lya halos seen around
LBGs could also explain our measurements.
 
\end{abstract}
\begin{keywords}
quasars: absorption lines -- galaxies: high-redshift -- galaxies:ISM -- galaxies: star formation
\end{keywords}

%\maketitle

\section{Introduction}
\label{sec:intro}

Damped \lya systems (DLAs) are the highest H~{\sc i} column density
absorbers seen in QSO spectra, with $N$(H~{\sc i}) $\ge$
2$\times$10$^{20}$\,cm$^{-2}$. These absorbers trace the bulk of the
neutral hydrogen at 2 $\le z \le$ 3
\citep{Prochaska2009ApJ...696.1543P,Noterdaeme2009A&A...505.1087N,Noterdaeme2012A&A...547L...1N}
and have long been considered to arise from the high-redshift
precursors of present day galaxies \citep[for a review
  see,][]{Wolfe2005ARA&A..43..861W}. Presence of enriched elements
\citep{Pettini1994ApJ...426...79P}, measured excitation of C~{\sc ii}
fine-structure levels
\citep{Wolfe2003ApJ...593..215W,Srianand2005MNRAS.362..549S},
existence of a correlation between metallicity and the velocity spread
of low-ion absorption lines \citep{Ledoux2006A&A...457...71L} akin to
the mass metallicity relation seen in galaxies and rotational
excitation of high $J$ level of H$_2$ detected in a small fraction of
DLAs
\citep{Ge97,Ledoux2003MNRAS.346..209L,Noterdaeme2008A&A...481..327N}
etc., suggest DLAs are associated in some way with star forming
regions. Even if one associates a moderate star formation rate (SFR)
to DLAs, they will contribute appreciably to the global inventory of
the cosmic star formation rate density at high-$z$
\citep{Wolfe2003ApJ...593..235W, Srianand2005MNRAS.362..549S}.

 A straightforward way to establish the link between the \hi gas and
 stellar components in DLAs is to detect the DLA host galaxies in line
 or continuum emission. A wide variety of galaxies hosting DLAs are
 found in imaging studies at low-$z$ \citep[i.e.,
   $z\le1$;][]{Chen2003ApJ...597..706C, Rao2003ApJ...595...94R}.
 However, till date only a handful of detections of galaxy
 counterparts hosting DLAs at $z > 2$ are confirmed using spectroscopy
 \citep[][]{Moller2004A&A...422L..33M,Fynb2010MNRAS.408.2128F,
   Peroux2011MNRAS.410.2251P,Bouche2012MNRAS.419....2B,Fynbo2010MNRAS.408.2128F,Noterdaeme2012A&A...540A..63N,Krogager2012MNRAS.424L...1K,Krogager2013MNRAS.433.3091K,
   Jorgenson2014ApJ...785...16J,Kashikawa2014ApJ...780..116K,Rubin2015ApJ...808...38R,Hartoog2015MNRAS.447.2738H,Srianand2016MNRAS.460..634S}.
 Based on all these efforts, the DLAs with emission line detections
 are found to trace a galaxy population with SFR of $\rm 0.4 - 25~
 M_\odot~ yr^{-1}$ and impact parameters ranging from 0.4$-$182 kpc,
 with a mean value of $\sim$ 25
 kpc~\citep{Krogager2012MNRAS.424L...1K,Peroux2011MNRAS.410.2251P}.
 \citet{Christensen2014MNRAS.445..225C} measured stellar masses of
 these DLA host galaxies. These are found to be consistent with the
 expected values based on mass-metallicity relations of high-$z$ galaxies. Note, the number
 of such DLAs are much smaller than the known number of DLAs or
 galaxies with spectroscopic redshifts. This is mainly because it is
 found to be challenging as the glare of the bright background quasar
 makes it difficult to detect the faint galaxy producing the DLA
 absorption at small impact parameters
 (\citealt{Lowenthal1995ApJ...451..484L,Bunker1999MNRAS.309..875B,Kulkarni2000ApJ...536...36K},
 and several unreported attempts).

Issues related to detecting faint galaxies against the glare of a
bright quasar can be easily addressed if one can find sightlines with
two or more optically thick \hi absorbers
\citep{Steide1992AJ....104..941S,
  OMeara2006ApJ...642L...9O,Christensen2009A&A...505.1007C,Fumagalli2010MNRAS.408..362F}.
In these cases the continuum emission from the foreground DLA can be
detected at wavelengths less than 912~\AA\ in the rest frame of the
higher redshift optically thick \hi absorber that removes the quasar
light completely. Using this so called the ``Double-DLA'' technique
\citet{Fumagalli2015MNRAS.446.3178F} have placed a stringent
constraints on the \emph{in-situ} star formation rates, SFR
$<0.09-0.27$ M$_\odot$ yr$^{-1}$ at the position of the absorbing gas.
By looking at the galaxies detected around the QSO sightlines they
concluded that most of the DLAs do not originate from highly
star-forming galaxies that are coincident with the absorbers. \par

The effective optical radius of high-$z$ galaxies are small
\citep[i.e., $\sim$ 1 kpc,][]{Shibuya2015ApJS..219...15S}. If we
conjecture that the DLAs are associated to such galaxies then the
observed number of DLAs per unit redshift requires \hi gas to be
extended appreciably beyond the stellar regions. Image stacking
analysis of high-$z$ galaxies suggests that the stellar distribution is also extended
\citep{Hathi2008AJ....135..156H,Rafelski2011ApJ...736...48R}. By
comparing the covering factor of diffuse emission from Lyman Break
Galaxies (LBGs) and \hi gas in DLAs (with an assumed star formation
efficiency) \citet{Rafelski2011ApJ...736...48R} argued that the star
formation in the outskirts of galaxies (i.e., $\sim$ 6 kpc from the
LBG's core) may be dominated by atomic gas probed by DLAs with
log~\nhi $\ge 21$ but with at least a factor 10 less efficiency
compared to that seen in local stellar disks. If the framework
presented by \citet{Rafelski2011ApJ...736...48R} is true then in a DLA
sample of log~\nhi $\ge 21$ one should see evidence for low surface
star formation along the line-of-sight and a faint galaxy within
$\sim$ 6 kpc. Similarly narrow band image stacking analysis of LBGs at
the rest frame \lya wavelength have shown diffuse \lya emission
extending up to several tens of kpcs
\citep{Steidel2011ApJ...736..160S}. In addition,
\citet{Rauch2008ApJ...681..856R} have detected a population of faint
\lya emitters with spatially extended \lya emission that have a total
cross-section consistent with that of DLAs. While the \lya scattering
requires only \nhi $\ge 2\times 10^{17}\ \rm cm^{-2}$ to generate
large diffuse \lya emission around LBGs, associating these scattering
regions in the LBG's outskirts to DLAs will also mean that the low
impact parameter regions [that will contribute to the high
    \nhi\ systems \citep[see,][]{Krogager2012MNRAS.424L...1K}] can
have associated \lya emission even if there is no \emph{in-situ} star
formation being present. All this suggests that the spectral stacking
analysis will be very useful in probing DLA-galaxy connection. \par

Interestingly, in spectroscopic surveys using fibers (e.g., 3 and 2
arcsec diameter fibers employed in SDSS-DR7 and SDSS-DR12 (BOSS),
respectively) one integrates, in addition to the light from a distant
quasar, the light from all the fore-ground galaxies that happen to
fall within the fiber along our line-of-sight. This allows detection
of nebular emission lines from such galaxies on top of the QSO
spectrum \citep{Wild2007MNRAS.374..292W,
  Noterdaeme2010MNRAS.403..906N,Menard2011MNRAS.417..801M,Straka2015MNRAS.447.3856S}.
In this regard, the \lya emission holds great potential in determining
the nature of high-$z$ DLA hosts. Since the direct detections of \lya
emission from high-$z$ galaxies hosting DLAs are very rare, several
studies have attempted to detect the \lya emission in the composite
spectra
\citep{Rahmani2010MNRAS.409L..59R,Noterdaeme2014A&A...566A..24N}. For
instance, in a stacking analysis of 341 DLAs of mean redshift $z
\sim\ $2.86, and log~\nhi $\ge 20.62$ seen in the SDSS quasar spectra,
\citet{Rahmani2010MNRAS.409L..59R} have found a $3 \sigma$ upper limit
on SFR of $\rm \le 1.2\ M_\odot~ yr^{-1}$. Using $\sim$ 95 extremely
strong DLAs (ESDLAs) with log~\nhi $\ge 21.7$ in the SDSS-DR12 spectra
\citet{Noterdaeme2014A&A...566A..24N} have detected the \lya emission
with an average luminosity of $\rm 6 \times 10^{41} erg\ s^{-1}$ in
the DLA core. These contemporary studies motivate us to unveil the
average SFR of DLAs in the stacked spectra by exploiting the
unprecedented number of DLAs found in the recent SDSS-BOSS survey
\citep[][]{Noterdaeme2009A&A...505.1087N,Noterdaeme2012A&A...547L...1N}.
Note that, this compilation contains $\sim 14972$ DLA systems with
log~\nhi $\ge 20.3$ which is a factor of 2.2 larger than the DLA
compilation based on SDSS-DR9 by
\citet{Noterdaeme2012A&A...547L...1N}. The available information from
emission and absorption stacks will not only allow us to detect the
\lya emission but also to probe its dependence on various quantities,
such as H~{\sc i} column density, rest equivalent width of metal
absorption lines, QSO colors, and absorption redshift.

This article is organized as follows. Section 2 describes our sample
selection criteria. In Section 3, we present details of our analysis
and average \lya emission for our full sample and various sub-samples.
Section 4 presents discussion based on absorption lines detected in
different stacked spectra. In Section 5, we discuss our results in the
framework of different possible scenarios. A detailed summary of our
study is presented in Section 6. Throughout, we have assumed the flat
cosmology with $H_0 = \rm 70\ km~ s^{-1}~ Mpc^{-1}$, $\rm \Omega_m =
0.3$ and $\rm \Omega_{\Lambda} = 0.7$.

\section{Sample}
\label{lab:sample}

 \begin{table}
 \centering
 \begin{minipage}{120mm}
 {\scriptsize
 \caption{ Details of our DLA sample.}
 \label{tab:sample}
 \begin{tabular}{@{} l l r @{}}
 \hline 
 \multicolumn{1}{c}{Sample }   &\multicolumn{1}{c}{Criteria}  &  \multicolumn{1}{c}{DLA systems } \\
\hline
Total DLAs                       & log \nhi$ \ge 20.3 $           & 14515  \\    \\

Primary sample                   & $ 2.3 \le$ \zabs $\le 3.44  $    & \\
                                 & $ CNR \ge 5 $                  & \\ 
                                 & log \nhi$ \ge 21.0 $           & 830 \\ 
                                 & $\beta \ge $5000~\kms        & \\ 
                                 & non-BAL                      & \\ 
                                 & \zabs $ > z$(\lyb)         & \\   \hline
                                 &                   & \\ 
New Measurements                 & log $N$(H~{\sc i}) $< 21.0  $        & 62 \\
                                 & Noisy Spectra             & 8 \\
                                 & Uncertain z               & 52 \\
                                 & Uncertain \nhi            & 4 \\  \hline
                                 &                   & \\ 
Final sample                     &                   & 704 \\ 
 \hline                                                                                 
 \end{tabular} 
 }                                                   
 \end{minipage}
 \end{table}

For our analysis, we have constructed a sample of systems from the
BOSS-DR12 DLA catalog, based on automatic search for DLAs in the
quasar spectra \citep[see,][]{Noterdaeme2012A&A...547L...1N} in
SDSS-III BOSS Data
Release{\footnote{\href{https://www.sdss3.org/}{https://www.sdss3.org/}}}.
We use the following stringent criterion: (1) we consider only systems
with log~\nhi $\rm \ge 21.0$. In this case the \lya absorption has a
dark core (with $\tau \ge 10$) spread at least over 7 times the
average FWHM ($\sim 160\ \rm km~s^{-1}$) of the instrumental profile
of the BOSS spectrograph. (2) We consider DLAs detected in spectra
with a median continuum-to-noise ratio (CNR) $\ge 5$. This cut is used
to ensure an accurate determination of the \hi column density. The CNR
is defined over a redshift range with the minimum redshift set redward
of any possible Lyman break present in the spectrum and the maximum
redshift defined at $5000$~\kms\ blueward of QSO emission redshift
\citep[see also ][]{Noterdaeme2009A&A...505.1087N}. (3) We consider
only DLAs that are in the redshift range $ \rm 2.3 \le$ \zabs $\le
3.44 $. This redshift range is chosen to avoid the poor signal-to-noise regions
in the blue spectrum and to exclude wavelength ranges affected by the
residuals from the poorly-subtracted sky emission lines in the red.
(4) We restrict ourselves to the observed wavelength range redward of
the \lyb $+$ \ovi emission line of the quasar to avoid any broad
associated \ovi absorption being misidentified as a DLA. (5) We avoid
DLAs located in the vicinity of the quasar (i.e., the so called
proximate DLAs) by considering the systems with velocity offset of
$\ge$ 5000~\kms\ with respect to the quasar emission redshift. After
applying these conditions and avoiding sightlines having broad
absorption lines from QSO outflows (i.e., BALQSOs), we have a sample
of 830 DLAs satisfying the above criterion.

To avoid any uncertainty in the absorption redshift and/or \hi column
density measurements, we further visually inspected the entire sample
and refitted the DLA absorption with a voigt profile. As accurate
  redshift measurement is critical for detecting the signal in the
  coadded spectrum, we measure the absorption redshift by cross-correlating the
  low-ionization metal absorption lines (e.g., C II$\lambda$1334, Si
  II$\lambda$1526, Al II$\lambda$1670, Fe II$\lambda$1608 , 2344,
  2374, 2382, 2586, 2600, MgII$\lambda$2796, 2803) redward to \lya
  emission. When no metal absorption line is detected we keep the
  redshift obtained from the \hi Lyman series transitions. This
  resulted in a final sample of 704 DLA systems with good \zabs\ and
  \nhi\ measurements. Among them, $\sim$80\% of the DLAs have redshift
  determined from metal lines. In the case of 62 DLAs the re-measured
$N$(H~{\sc i}) turns out to be less than our cutoff value 10$^{21}$
cm$^{-2}$, thereby they go out of our primary sample. Similarly for 64
cases visual inspections, independently by two of the co-authors of
this paper, suggest that the DLA trough is noisy and lack of higher
H~{\sc i} Lyman series lines and associated metal lines prevent
accurate measurements of $N$(H~{\sc i}) and $z$.  These 64 systems
  are also removed from our primary sample. Details of our sample are
summarized in Table~\ref{tab:sample}. \par

 \begin{figure}
   \epsfig{figure=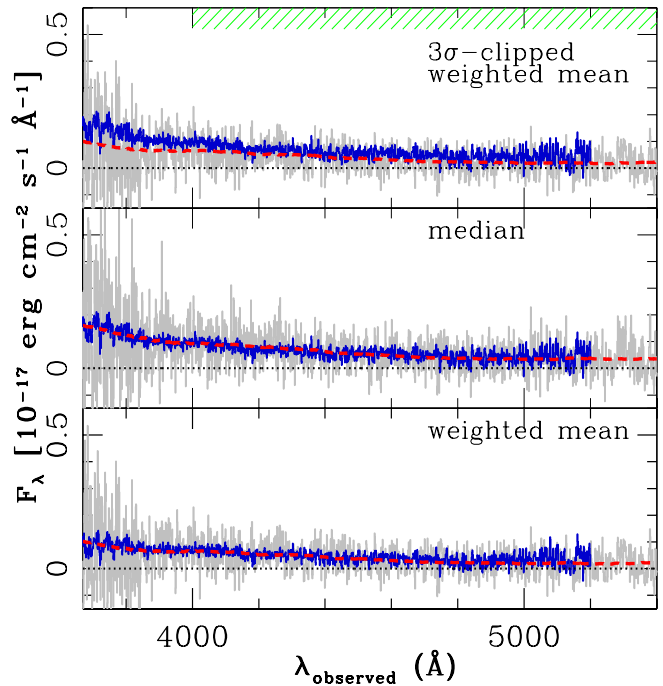,height=9.5cm,width=9.0cm}
 \caption{ The average residual flux in the Lyman limit region of DLAs
   (blue solid curve) and in the DLA core pixels with $\tau \ge 10$
   (gray curve) are shown as a  function of the observed wavelength. The
   dashed line represents the running average of residual flux in DLA
   core pixels, which is used as a background template for correcting
   the residuals in the DLA core. The shaded region in the top panel
   shows the wavelength range of interest for our stacking analysis.}
 \label{fig:DLACORE} 
   \end{figure}

While most of the results presented here are based on this sample,
sometime to test our results based on smaller sub-samples, in the
Appendix we also present result based on samples where we have
released the limiting CNR that defines our sample. The reduced
one-dimensional spectra for our entire sample of 704 quasars were
downloaded from the SDSS-BOSS Data Archive
Server{\footnote{\href{http://dr12.sdss3.org/bulkSpectra}{http://dr12.sdss3.org/bulkSpectra}}.
  Details about the BOSS spectral information can be found in
  \citet{Dawson2013AJ....145...10D}. Briefly, the BOSS spectra were
  obtained with a fiber having a diameter of 2 arcsec (a maximum
  impact parameter of $\sim$ 8~kpc for the above assumed cosmology)
  and cover a spectral range from $3650 - 10,400$~\AA\ at a resolution
  ($\lambda/\Delta \lambda$) of about $1560-2270$ in the blue (i.e.,
  $\lambda \le 6350$~\AA) and $1850-2659$ in the red channels.

\section{Analysis}
\label{lab:analysis}
\subsection{The composite spectra}
\label{subsec:compositespec}
We have generated several composite spectra with an aim to study both
the average emission and absorption properties of  high column
  density DLAs.  For stacking, an individual spectrum is shifted
  to the rest-frame of the DLA, while conserving the flux and
  rebinning onto a common grid, keeping the same pixel size (constant
  in velocity space) as the original data
  ~\citep{Bolton2012AJ....144..144B}. We produce the median,
continuum normalized stacked spectrum for studying the absorption
lines of DLAs. Each individual spectrum is normalized using the
best-fit principle component analysis (PCA) continuum model of
\citet{Bolton2012AJ....144..144B}. We apply a median-smoothing filter
to remove additional imperfections on intermediate scales in the PCA
continuum normalized spectra. For this, we mask out pixels possibly
containing narrow absorption lines, by only keeping fluctuations
within 1.5$\sigma$ of the continuum. We mainly consider regions
redward of the \lya emission, to avoid the blending by the \lya
forest. The normalized flux error in each pixel of the stacked
spectrum is the absolute deviation measured around the mean flux.
These spectra are used to study any possible correlation between
average absorption properties and the \lya emission.\par

For studying the \lya emission, we convert each spectrum into
luminosity per unit wavelength using the luminosity distance at the
redshift of the DLAs for the above mentioned cosmological parameters.
To remove contribution from any possible outliers, we have used the
weighted mean and the $3 \sigma$ clipped weighted mean statistics with
1/$\sigma^2$ weighting with $\sigma$ being the error in the
luminosity. However, we note that using $1/\sigma^2$ weighting will
give more weightage to the low redshift systems, when we use the rest
frame spectra in luminosity. Therefore, we have constructed the median
composite spectrum also. As the DLAs are characterized by a wide
  flat absorption trough with zero transmitted flux (i.e., dark core),
  we do not re-scale or normalize the spectrum before the co-addition.
  The $1\sigma$ uncertainty over each pixel in the stacked spectrum is
  estimated from the central interval encompassing 68\% of the flux
  distribution of the corresponding pixel. \par

\begin{figure}
  \epsfig{figure=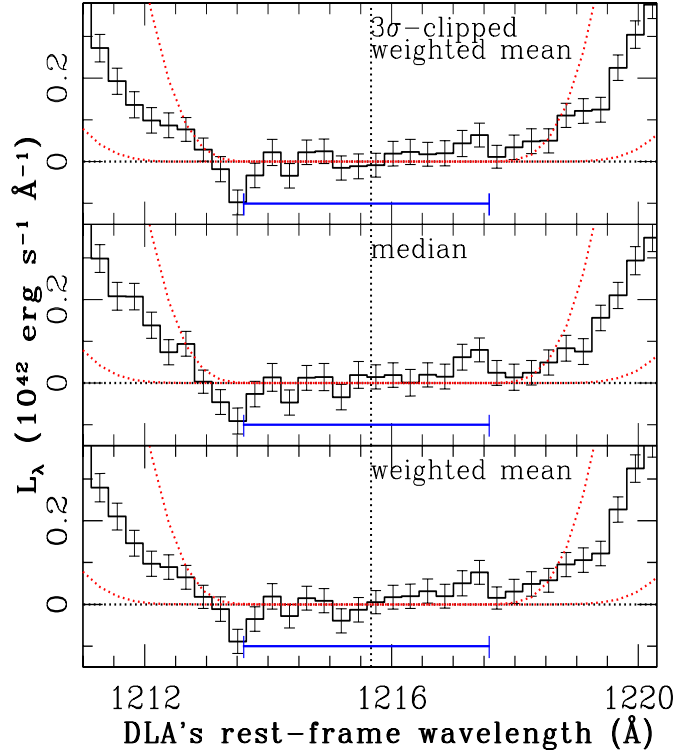, height=10.5cm,width=9cm} 
  \caption{ The stacked spectrum corresponds to the 3$\sigma$ clipped
    weighted mean (\emph{top}), median (\emph{middle}) and weighted
    mean (\emph{bottom}), for DLAs in the redshift range of $ 2.30 \le
    z \le 3.44$. The blue segment shows the DLA core with $\tau \ge
    10$ for log~\nhi\ $\rm = 21.0$. The dashed curves show
    the synthetic profiles for lower (i.e., log~\nhi $ = \rm
    21)$ and median (i.e., log~\nhi $\rm = 21.23)$
    column density of DLAs used to get the stacked spectrum. }
  \label{fig:stack_dla_flg0}
   \end{figure}

 To check and correct the non-zero flux offsets \citep[e.g., see
   also][]{Rahmani2010MNRAS.409L..59R,
   Cai2014ApJ...793..139C,Noterdaeme2014A&A...566A..24N} in the core
 of DLAs due to poor background subtraction we follow two methods: (i)
 The average residual flux is calculated by stacking the Lyman limit
 region (i.e., $\lambda \le 912\times (1+z)$ \AA) of DLAs in the
 observer's frame, where the average flux is expected to be negligibly
 small due to the large expected \hi optical depth. For this, we have
 used 3259 QSO spectra having DLAs in the redshift range $2.95\le z\le
 4.6$ whose Lyman limit falls in the observed wavelength range
 3600-5100~\AA. The stacked spectrum based on three different
 statistics discussed above are shown in three different panels in
 Fig.~\ref{fig:DLACORE} with a blue solid histogram. It is clear that
 there is a non-zero residual flux present over the redshift range
 $2.0\le z \le 3.2$ (i.e., $3650 \le \lambda \le 5100$~\AA) for the
 \lya absorption. This residual flux could arise from leakage of
 far-ultraviolet radiation from the quasar host galaxy through the DLA
 \citep{Cai2014ApJ...793..139C} or due to imperfect fiber sky
 subtraction \citep{Bolton2012AJ....144..144B} and should be corrected
 for before we do the stacking. However, for DLAs in our sample, at
 $3.2 \le z \le 3.4$, we do not have the better residual measurements
 from the Lyman limit systems. For this $z$ range we use the second
 method. (ii) We construct the residual flux template (background) by
 stacking the DLA core pixels with $ \tau \ge 10$ that fall within
 this observed wavelength range. To be specific, we have used the core
 pixels of all the available DLAs in addition to those that are part
 of our sample. We generated a template spectrum by considering only
 the DLA core pixels with optical depth $ \tau \ge 10$ for a given
 \nhi\ value. In Fig.~\ref{fig:DLACORE}, we also show the average flux
 in the stacked spectrum of DLA core regions, constructed using
 $\sim$7072 DLAs with log \nhi $ \ge 20.62$ and average CNR $\ge 2$
 (gray curve). We construct the background for all the three stacking
 statistics and found that in the overlapping observed wavelength
 range the residual flux in the Lyman limit region and core of the DLA
 absorption profile matches well. Besides, we note that the residual
 flux template does not depend on the brightness and color of the
 background quasar. Next, in order to account for this average
 residual flux, we compute its running average (dashed curve) over 300
 pixels and constructed the background template for each statistics.
 We find that the background template obtained by varying the
 smoothing window in the range of 100-300 pixels gives similar
 results. Finally, before generating the stacked spectrum in the DLA
 rest frame, we correct each spectrum for the non-zero flux offset by
 subtracting the flux given by the template for the corresponding
 statistics. \par

In addition, to the non-zero average flux in the bottom, we also
notice in several DLAs there are large pixel-to-pixel variations in
the core pixels (see few examples in Fig.~\ref{fig:bad_pix_appx} in
Appendix). While we could associate this with the residual from telluric
line subtraction for few cases, the origin of such fluctuations in
most cases is unclear. By and large these pixel-to-pixel variations do
not affect our results, but may have strong influence when we are
dealing with sub-samples having much smaller number of systems.  As
stated earlier in such cases we relax the CNR constraint that defines
our main sample to have additional systems to test the results we
get from our main sample.

\subsection{The average \lya emission of DLAs}
\label{subsec:avg_lya_profile}

 \begin{table*}
 \centering
 \begin{minipage}{150mm}
 {%\scriptsize
 \caption{ \small {The \lya line luminosity measured from the stacked
     spectra.}}
 \label{tab:line_lum_para}
 \begin{tabular}{@{} l c c c r r r@{}}
 \hline 
 \multicolumn{1}{c}{criteria } &\multicolumn{1}{c}{number} &\multicolumn{1}{c}{$\langle$logN{\sc (H i)}$\rangle$} &\multicolumn{1}{c}{$\rm \langle$\zabs $\rangle$} &\multicolumn{3}{c}{\lya luminosity ($\times 10^{40} \rm erg~s^{-1}$)}     \\
                         &  of   &     &&\multicolumn{1}{c}{weighted mean}  &  \multicolumn{1}{c}{3$\sigma$ clipped } &\multicolumn{1}{c}{median} \\
                         & systems   &     & & &\multicolumn{1}{c}{  weighted mean       } &\\
\hline

Final sample                      & 704   & 21.23 & 2.71 &    4.54 $\pm$   3.12  &    4.70 $\pm$    3.19  &    5.24  $\pm$   3.31  \\ 
 \nhi $ \ge 10^{21.23} ~cm^{-2}$   & 354   & 21.40 & 2.72 &    8.83 $\pm$   5.05  &    9.15 $\pm$    5.13  &    6.75  $\pm$   5.32  \\
 $10^{21} \le$ ~\nhi  $< 10^{21.23} \rm ~cm^{-2}$     & 350   & 21.10 & 2.69 &    3.06 $\pm$   4.32  &    3.27 $\pm$    4.36  &    6.29  $\pm$   4.65  \\
  \zabs  $\rm \ge 2.7         $   & 352   & 21.23 & 2.97 &    6.21 $\pm$   4.80  &    6.18 $\pm$     4.78 &    7.96  $\pm$   5.00  \\
  \zabs  $\rm < 2.7         $     & 352   & 21.22 & 2.48 &    3.41 $\pm$   4.12  &    4.60 $\pm$    4.35  &    3.84  $\pm$   4.39  \\
 $W_{1526} \ge 0.8$ \AA  {\textcolor{blue}{$^a$}}       & 288   & 21.25 & 2.68 &  10.51 $\pm$   4.95  &   13.62 $\pm$    5.02  &   15.82  $\pm$   5.18  \\
 $W_{1526} <   0.8$ \AA            & 406   & 21.26& 2.75 &   4.07 $\pm$   4.67  &   4.63 $\pm$    4.05  &    4.71   $\pm$   5.02  \\
($r-i$) $\ge 0.1$                 & 352   & 21.24 & 2.82 &   6.22  $\pm$   4.49  &    5.43  $\pm$   4.56  &   10.93  $\pm$   4.70  \\
($r-i$) $< 0.1$                   & 352   & 21.22 & 2.59 &   3.19  $\pm$   4.34  &    5.66  $\pm$   4.47  &    4.02  $\pm$   4.66  \\
 \hline                                                                       
 \end{tabular} 
 }
{ \textcolor{blue}{$^a$} {Only systems with  \siia\ absorption line detection are considered.}}                                               
 \end{minipage}
 \end{table*}

In Fig~\ref{fig:stack_dla_flg0}, we show our stacked spectrum of 704
DLA systems obtained using the weighted mean (\emph{bottom panel}),
median (\emph{middle panel}) and $3 \sigma$ clipped weighted mean
(\emph{top panel}) statistics. Our stacked spectra show an asymmetric
flux distribution in the bottom of the DLA absorption profile, with a
possible enhanced flux in the red half (i.e., $\lambda_r >
1215.67$~\AA) within the core region. Such asymmetric \lya line
profiles are naturally produced by the radiative transport of \lya
photons in an outflowing medium
\citep{Neufeld1990ApJ...350..216N,Neufeld1991ApJ...370L..85N,
  Verhamme2006A&A...460..397V, Hansen2006MNRAS.367..979H,
  Dijkstra2006ApJ...649...14D,Dijkstra2014PASA...31...40D}. \par

It is not a straight forward exercise to quantify the Ly$\alpha$ line
luminosity (or its upper limit) as the resonant scattering of
Ly$\alpha$ photons along with the gas kinematics introduces a wide
range of velocity offsets for the escaping Ly$\alpha$ photons. In
addition, by stacking, we average over a large number of systems with
varied \lya optical depths and velocity fields. Therefore, we estimate
three \lya luminosities from each stacked spectrum: (i) integrated
luminosity over the entire DLA core region, (ii) integrated luminosity
($\rm L^{b}_{\lambda}$) in the blue part (i.e., core regions having
$\lambda_r < 1215$~\AA) and (iii) the same ($\rm L^{r}_{\lambda}$) in
the red part (i.e., $\lambda_r > 1215$~\AA). We define the core region
of the DLA absorption profile to be the wavelength range where the
\lya optical depth, $ \tau \ge 10$, for the minimum column density of
log~[\nhi$ ~\rm / cm^{-2}]\sim$21 in our sample (see above in
Section~\ref{lab:sample}). This is indicated in each panel as a
horizontal line. The integrated \lya luminosity over the DLA core
regions are summarized in Table~\ref{tab:line_lum_para}. For the full
sample, we do not detect any statistically significant signal above
the 2$\sigma$ level. All three statistics give nearly same integrated
luminosities and errors suggesting that the stacked spectrum is not
severely influenced by few outliers. \par

\begin{figure}
\epsfig{figure=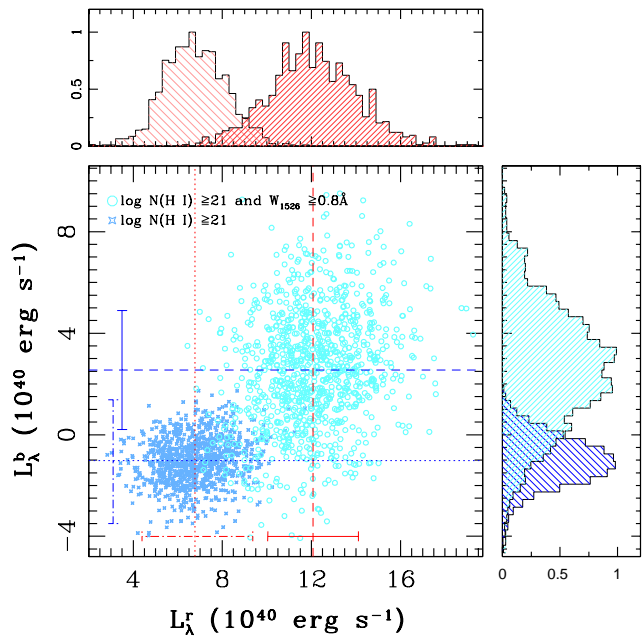,height=8.5cm,width=8.5cm}
  \caption{ Distribution of the measured mean \lya luminosities on the
    red (${\rm L_\lambda^r}$) and blue (${\rm L_\lambda^b}$) part of
    the DLA core region in the bootstrapped stacked spectra for our
    full sample (\emph{blue star}) and for the sub-sample with $\rm
    W_{1526} \ge 0.8~\AA$ (\emph{cyan open circle}), using $\sim$1000
    random sub-samples generated by considering only 90\% size of the
    sample. The vertical and horizontal (\emph{dashed and dotted})
    lines represent the mean bootstrapped luminosity in the red and
    blue part.  The bar shows $1\sigma$ deviation of the bootstrap
      sample around the mean. No correlation is seen between the $\rm
    L^{r}_{\lambda}$ and $L^{b}_{\lambda}$, with a Spearman's rank
    correlation coefficient ($r_s$) of 0.16 and 0.19, respectively for
    both the sub-samples. We also show the histogram distribution of
    $\rm L^{r}_{\lambda} and\ L^{b}_{\lambda}$ in the figure.}
  \label{fig:boot_lya}
   \end{figure}

  Furthermore, to account for the uncertainties related to under/over
  subtraction of non-zero flux (i.e. background correction) and to
  test if the stack is influenced by few outliers, we perform a
  bootstrap analysis. For this, we make stacked spectra of 1000
  sub-samples by randomly selecting 90 and 80 per cent of the sample.
  Note that as we have performed sub-sampling bootstrap with 90\% of
  the sample, the measured statistical error from bootstrap sample is
  smaller than the uncertainties in each realization (given by the
  size of the interval including 68\% of the pixel value). Therefore,
  the uncertainty over each pixel is taken to be the maximum error
  value over 1000 realization.

  In this exercise, we also test the presence of any systematic
  differences between the integrated flux in the blue and red part of
  the DLA core in the stacked spectra. Results of this bootstrap
  analysis are summarized in Table~\ref{tab:line_para}. Here we
  consider only the median stacked and 3$\sigma$ clipped weighted mean
  spectrum. It is clear from this table that the total integrated
  luminosity over the core region of the DLA absorption trough are
  $(5.8\pm3.5)\times 10^{40}~{\rm erg~s^{-1}}$ and $(5.3\pm3.7)\times
  10^{40}~{\rm erg~s^{-1}}$ respectively when we consider sub-samples
  with 90 per cent and 80 per cent of the total sample and median
  stacked spectrum. In Table~\ref{tab:line_para}, we also give the
  \lya luminosity integrated over the blue and red parts alone. It is
  clear that, while the luminosity is consistent with zero in the blue
  part [i.e., ${ \rm L}_{\lambda}^b = (-1.0\pm2.4)\times 10^{40}~{\rm
      erg~s^{-1}}$], there is a 2.8$\sigma$ excess seen in the red
  part [i.e., ${ \rm L}_{\lambda}^r = (6.9\pm2.5)\times 10^{40}~{\rm
      erg~s^{-1}}$]. The distribution of the measured luminosities in
  the blue and red part in individual median stacked spectrum
  considering 90 per cent of the total sample in each sub-sample is
  shown with stars in Fig~\ref{fig:boot_lya}. First of all, we note
  that there is no correlation seen between the \lya luminosity in red
  and blue part, with a Spearman's rank correlation coefficient
  ($r_s$) of 0.16. It is also clear from the figure that the average
  \lya luminosity of the blue part is consistent with zero, whereas,
  the red part always shows a positive luminosity. From
  Table~\ref{tab:line_para}, we also see this trend being present even
  when we do bootstrap analysis for 3$\sigma$ clipped weighted mean
  spectra albeit with slightly lower significance level. {\it The main
    conclusion from this section is that the average \lya luminosity
    of DLAs with log~$N$(H~{\sc i}) $\ge$21 within 8 kpc to the QSO
    sightline, is less than one hundredth of the \lya luminosity of an
    $\rm L_{\star}$ galaxy (i.e., $\rm L_{\star}$(Ly$\alpha$) $= 5.8\ \times
    10^{42}\ erg\ s^{-1}$) at these redshifts
    \citep[][]{Ouchi2008ApJS..176..301O}. There is a hint of peak
    emission in the \lya profile being redshifted with respect to the
    DLA systemic redshift.}

In what follows, we will explore the dependence of \lya luminosity on
different measurable parameters of the DLA.

\subsection{Dependence on $N$(H~{\sc i}):}

The H~{\sc i} column density of DLAs is found to be strongly
anti-correlated with the impact parameter of the quasar sightline,
where a larger \nhi\ typically originates from a sightline with
smaller impact parameter
\citep{Zwaan2005MNRAS.364.1467Z,Rao2011MNRAS.416.1215R,Peroux2011MNRAS.410.2251P,Krogager2012MNRAS.424L...1K}.
In this scenario, we expect the galaxies associated with  high
H~{\sc i} column density systems to preferably come inside the fiber
whereas those associated with  low column density will contribute
less to the \lya emission in the fiber spectrum. Recall, that the
fiber radius of 1 arcsec corresponds to a physical size of $\sim 8 \rm
\ kpc$ at a median $z$ of 2.7. From figure 8 of
\citet{Zwaan2005MNRAS.364.1467Z}, we see that a line-of-sight having
log~\nhi$\ge21$ will occur, at an impact parameter $\le $ 8~kpc, 75
per cent of the times. Whereas, the same for log~\nhi$\ge21.7$ will
occur within an impact parameter of 8~kpc in 90 percent of the cases.
However, results for high-$z$ DLAs suggest that a given $N$(H~{\sc i})
could come from a larger impact parameter
\citep[see][]{Rao2011MNRAS.416.1215R,Peroux2011MNRAS.410.2251P} than
the one discussed by \citet{Zwaan2005MNRAS.364.1467Z} based on H~{\sc
  i} emission from  low-$z$ galaxies.

On the other hand, \citet{Toribio2011ApJ...732...93T} have found that
the size of the \hi emitting region in low-$z$ galaxies is
proportional to their optical luminosity. Because of this, it is
possible that a given \hi column density will originate from a large
impact parameter for a high luminosity galaxy compared to a low
luminosity one. It is also clear from figure 18 in
\citet{Zwaan2005MNRAS.364.1467Z} that when we have log~\nhi$\ge21$, at
an impact parameter of 10 kpc, the associated galaxy is most likely to
be a sub-L$_{\star}$ galaxy. \citet{Patra2013MNRAS.429.1596P}, using H~{\sc i}
maps of low-$z$ dwarf galaxies found roughly 10 per cent of the
cross-section above log~$N$(H~{\sc i}) = 20.3 at $z$ = 0 is provided
by dwarf galaxies. However, this fraction sharply falls to $\le 1$ per
cent by log~$N$(H~{\sc i}) $\sim$ 21.5 as the cross-section of such
gas in dwarf galaxies are very small. Typically these high column
density gas are located close to star forming regions even when there
is no one to one correspondence between them
\citep{Roychowdhury2014MNRAS.445.1392R}. In this scenario, for our
sample with log~\nhi\ $\ge$21, it is most likely that the stellar
light that will directly go through the fiber will come from less
luminous galaxies. The brighter galaxies contributing to the
  absorption likely fall outside the BOSS
  fiber~\citep{Lopez2012MNRAS.419.3553L}.

\begin{table*}
 \centering
 \begin{minipage}{190mm}
 {\small
 \caption{ \small{Luminosity of \lya line in bootstrap analysis for
     both the median and $3\sigma$ clipped weighted mean stacked
     spectra.}}
 \label{tab:line_para}
 \begin{tabular}{@{} l r r r r r r c c@{}}
 \hline 
 \multicolumn{1}{c}{Sample}   &\multicolumn{3}{c}{\lya luminosity ($\times 10^{40} \rm erg~s^{-1}$)} &\multicolumn{3}{c}{ \lya luminosity ($\times 10^{40} \rm erg~s^{-1}$)} \\
 &\multicolumn{3}{c}{median stack}&\multicolumn{3}{c}{$3\sigma \rm clipped\ weighted\ mean$} \\
 
                              & DLA-bottom  & \multicolumn{1}{c}{L$_\lambda^b$}   &   \multicolumn{1}{c}{L$_\lambda^r$}       &  DLA-bottom &      \multicolumn{1}{c}{L$_\lambda^b$}         &    \multicolumn{1}{c}{L$_\lambda^r$} \\
      
\hline
Full with 90\% bootstrap                       &   5.82 $\pm$  3.48{\textcolor{blue}{$^a$}} &    -1.06 $\pm$    2.44 &    6.88 $\pm$   2.49 &     $ 3.87\pm3.20 $ & $-2.27\pm 2.21  $&$6.14 \pm 2.32 $ \\
Full with 80\% bootstrap                       &   5.27 $\pm$  3.69 &    -1.32 $\pm$    2.58 &    6.58 $\pm$   2.65 &     $3.86 \pm 3.40$ & $-2.22\pm 2.35  $&$6.08 \pm 2.46 $ \\
\nhi $\rm \ge 10^{21.23} ~cm^{-2} $            &   6.24 $\pm$  4.95 &    -2.33 $\pm$    3.46 &    8.57 $\pm$   3.53 &      $  5.58\pm  4.62$&$ -2.51\pm  3.20 $&$ 8.09\pm  3.33$ \\
$10^{21} \le$ ~\nhi $<  10^{21.23} ~cm^{-2}$    &   3.55 $\pm$  4.82 &    -1.00 $\pm$    3.37 &    4.56 $\pm$   3.45 &      $  2.97\pm  4.45$&$ -1.72\pm  3.06 $&$ 4.69\pm  3.23$ \\
\zabs  $\rm \ge 2.7$                           &   7.34 $\pm$  5.28 &     1.72 $\pm$    3.68 &    5.61 $\pm$   3.79 &     $  6.03\pm  4.92$&$  1.17\pm  3.39 $&$ 4.86\pm  3.57$ \\
\zabs  $\rm <   2.7$                           &   4.02 $\pm$  4.60 &    -3.11 $\pm$    3.22 &    7.13 $\pm$   3.29 &     $  2.66\pm  4.27$&$ -4.64\pm  2.96 $&$ 7.30\pm  3.07$ \\
$W_{1526} \ge 0.8$ \AA{\textcolor{blue}{$^\dagger$}}                        &  14.55 $\pm$  5.50 &     2.49 $\pm$    3.87 &   12.06 $\pm$   3.91 &      $ 11.33\pm  5.06$&$  2.95\pm  3.51 $&$ 8.39\pm  3.64$ \\
$W_{1526}  <     0.8$ \AA                      &   0.64 $\pm$  4.55 &    -3.66 $\pm$    3.15 &    4.31 $\pm$   3.21 &      $  0.80\pm  4.43$&$ -5.20\pm  3.04 $&$ 5.28\pm  3.22$ \\
$W_{1526} \ge 0.4$ \AA{\textcolor{blue}{$^\dagger$}}                         &  13.49 $\pm$  4.05 &     1.16 $\pm$    2.84 &   12.34 $\pm$   2.90 &      $ 11.34\pm  3.74$&$ -0.79\pm  2.60 $&$12.13\pm  2.69$ \\
($r-i$) $\ge 0.1$                              &   9.32 $\pm$  4.97 &    -0.45 $\pm$    3.56 &    9.78 $\pm$   3.48 &      $  4.84\pm  4.59$&$ -2.57\pm  3.24 $&$ 7.41\pm  3.24$ \\
($r-i$) $< 0.1$                                &   3.45 $\pm$  4.93 &    -0.75 $\pm$    3.36 &    4.19 $\pm$   3.61 &      $  3.42\pm  4.48$&$ -1.59\pm  3.04 $&$ 5.02\pm  3.30$ \\
\hline                                                   
 \end{tabular} 
 } \\ { \textcolor{blue}{$^a$} {The individual errors are the maximum
     error seen in the 1000 realization during the bootstrap analysis
     and  not the statistical\\  error of the bootstrap sample.}} \\ 
 {
   \textcolor{blue}{$\dagger$} {Only systems with \siia\ absorption line
     detection are considered.}}\\

 \end{minipage}
 \end{table*}

All this motivates us to explore the \lya emission as a function of
the H~{\sc i} column density. In order to understand the dependence of
\lya luminosity on \nhi, we divided the sample into two around
log~\nhi\ = 21.23, the median log~\nhi\ of our sample. The median
log~\nhi\ of the two sub-samples are 21.1 and 21.4 respectively. The
\lya luminosities measured by integrating over the DLA core region are
given in Table~\ref{tab:line_lum_para}. For the weighted mean and
3$\sigma$ clipped weighted mean the \lya luminosity measured in the
high \nhi\ sub-sample is slightly higher than those measured from the
low column density sub-sample. However, the difference is not
statistically significant considering the associated errors in both
measurements. We also note that the measurements based on the median
stacked spectrum is similar for both sub-samples. We then consider the
bootstrapping results presented in Table~\ref{tab:line_para}. As in
the case of the whole sample, the \lya luminosities measured for the
red part are always higher than those measured for the blue part.
However, even for the red part the difference between two sub-samples
based on \nhi\ (i.e., $(4.0\pm4.9)\times 10^{40}~{\rm erg~s^{-1}}$ for
the median stacked spectrum) are not statistically significant. 
  However, we wish to note here that the range in \nhi\ is very small
  in our sample to reveal any weak trend in the \lya emission with
  \nhi. \par

\subsection{Dependence on $z$:}

 Next, we consider two sub-samples based on $z_{abs}$ by dividing the
 full sample at $z_{abs} = 2.7$. The sub-samples have median redshift
 of 2.48 and 2.97. From Tables~\ref{tab:line_lum_para}, we note that
 the \lya luminosity measured over the core region appears to be
 slightly higher for the high $z$ sample but the difference is not
 statistically significant. Table \ref{tab:line_para}, also suggests
 that there is no statistically significant difference in the average
 \lya luminosity as a function of $z$. As the projected size of the
 fiber in the whole $z$ range, over which our DLAs are detected, is
 nearly the same, it does not influence the $z$ evolution. However, in all
 sub-samples the \lya luminosity inferred is higher for the red part
 compared to that of the blue part, albeit with less statistical
 significance, compared to what we find for the full sample. We note
 that the smaller $z$ range in our sample provides a limited leverage
 to reveal any weak trend of \lya emission with $z$.

\begin{figure}

 \epsfig{figure=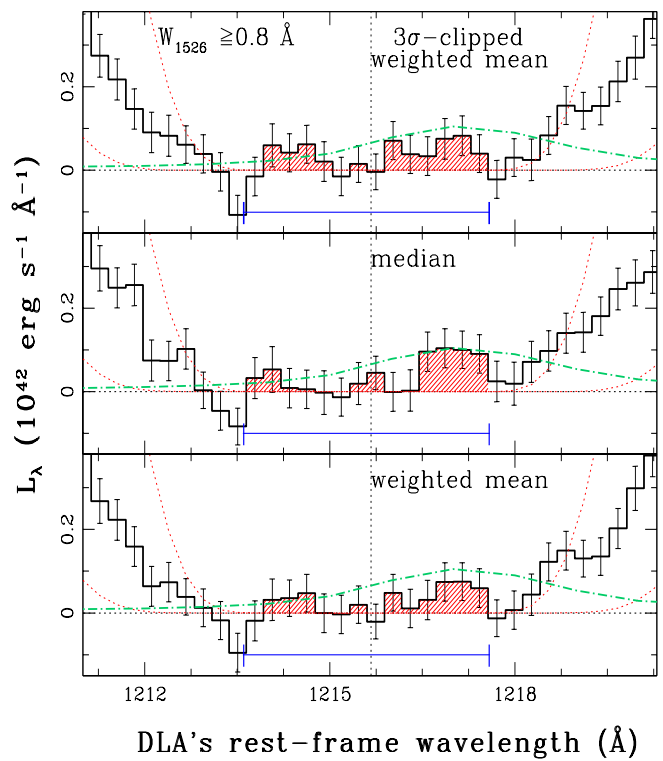, height=10.5cm,width=9cm} 
   \caption{Same as
     Fig.~\ref{fig:stack_dla_flg0}, for the sub-sample with $W_{1526}
     \ge 0.8$ \AA. The dashed curve is \lya emission in the composite
     spectrum of LBGs \citep{Shapley2003ApJ...588...65S} with its
     amplitude scaled to match the luminosity seen in our stacked
     spectrum.}
  \label{fig:stack_siew}
   \end{figure}

\subsection{Dependence on W$_{\rm r}$(Si~{\sc ii}$\lambda$1526)}
  
\citet{Prochaska2008ApJ...672...59P} found a strong correlation
between the rest equivalent width of Si~{\sc ii}$\lambda$1526 line
(denoted as $W_{1526}$) and metallicity \citep[see
  also][]{Jorgenson2013MNRAS.435..482J,Neeleman2013ApJ...769...54N}.
They found this correlation to be stronger than the correlation found
between low ion velocity widths ($\Delta v_{90}$) and metallicity
\citep{Ledoux2006A&A...457...71L}. The correlation between $W_{1526}$
and metallicity has also been confirmed by
\citet{Kaplan2010PASP..122..619K} in the case of metal strong DLAs.
The measured correlation implies an average metallicity of Z = $-
1.48$ Z$_\odot$ (respectively $-$1.06~Z$_\odot$) for $W_{1526}$ = 0.4
\AA\ (respectively 0.8 \AA) with a scatter of 0.25 dex in the
metallicity. The average metallicity will be 0.15 dex higher if we use
the coefficients found by \citet[][]{Jorgenson2013MNRAS.435..482J}. As
Si~{\sc ii}$\lambda$1526 is a strong transition, the absorption line
profile will be highly saturated for the high equivalent width
systems. Therefore, the measured equivalent width usually samples the
velocity field probed by the optically thin gas that is either part of
an outflow or in the halo. Thus, this correlation was interpreted as a
reflection of the underlying mass-metallicity relation in galaxies
\citep{Ledoux2006A&A...457...71L,Christensen2014MNRAS.445..225C}. \par

We find the median $W_{1526} =0.8$~\AA, if we consider DLAs with a
clear detection of \siia\ absorption line. Therefore, we consider two
sub-samples of DLAs with $W_{1526} < 0.8$~\AA\ and $\ge$0.8~\AA. The
median $W_{1526}$ of these two sub-samples are 0.53~\AA\ and 1.18~\AA.
The DLAs without Si{~\sc ii}$\lambda1526$ absorption line detections
are also included in the $W_{1526} < 0.8$~\AA\ sub-sample. From
Table~\ref{tab:line_lum_para} (and Fig.~\ref{fig:stack_siew}) we can
see that when we consider only systems with $W_{1526}\ge0.8$~\AA\ the
integrated \lya luminosity over the full DLA core itself is
significantly more from zero by more than 3$\sigma$ level (see
Table~\ref{tab:line_lum_para}). We find L =($15.8 \pm 5.2$) $\times
10^{40} \rm erg\ s^{-1}$, a detection at 3$\sigma$ level, in the
median stacked spectrum. The weighted mean and 3$\sigma$ clipped
weighted mean stacked spectra also show consistent results albeit with
slightly less significance. The measured \lya luminosity in this case
corresponds to 2 to 2.5\% of the \lya luminosity of an L$_{\star}$ galaxy at
these redshifts.
However, for the sub-sample of systems having $W_{1526} < 0.8$~\AA, we
find L ($\backsimeq 4.7 \pm 5.0$) $\times 10^{40} \rm erg\ s^{-1}$ in
the median stacked spectrum. Consistently lower luminosities are also
seen in the case of weighted mean and 3$\sigma$ clipped weighted mean
stacked spectra. The difference in the \lya luminosities between high
and low equivalent width sub-samples is = ($11.1 \pm 7.2$)$\times
10^{40} \rm erg\ s^{-1}$ for the median stacked spectrum. Therefore,
we have a marginal evidence for the \lya emission being high for high
\sii\ equivalent width systems.

Next, we consider $\rm L^{b}_{\lambda}$ and $\rm L^{r}_{\lambda}$ for
the two sub-samples. For the $W_{1526} \ge 0.8$~\AA\ sub-sample we
find the $\rm L^{r}_{\lambda}$ $\sim$ ($12.1 \pm 3.9$) $\times 10^{40}
\rm erg\ s^{-1}$ is higher than $\rm L^{b}_{\lambda}$ $\sim$ ($2.5 \pm
3.9$) $\times 10^{40} \rm erg\ s^{-1}$ in the median stacked spectrum.
The difference, ($9.6 \pm 5.5$)$\times 10^{40} \rm erg\ s^{-1}$, is
significant at a 1.7$\sigma$ level. The result of bootstrap analysis
retaining 90\% of the data is also shown in Fig.~\ref{fig:boot_lya}.
First of all we notice, unlike the full sample, the scatter is much
higher in the $W_{1526}$ based sub-sample. However, it is clear that
while there is no statistically significant detection of emission in
the blue part, in the red part non-zero fluxes are detected at more
than 3$\sigma$ level. Interestingly, even in the $W_{1526} <
0.8$~\AA\ sub-sample $\rm L^{r}_{\lambda} =$ ($4.3 \pm 3.2$) $\times
10^{40} \rm erg\ s^{-1}$ is stronger than $\rm L^{b}_{\lambda} =$
($-3.7 \pm 3.2$) $\times 10^{40} \rm erg\ s^{-1}$ in the median
stacked spectrum. We notice $\rm L^{r}_{\lambda}$ in the case of
$W_{1526} \ge 0.8$~\AA\ is higher than $\rm L^{r}_{\lambda}$ measured
for $W_{1526} < 0.8$~\AA\ sub-sample with the difference being ($5.1
\pm 5.5$) $\times 10^{40} \rm erg\ s^{-1}$. As can be seen from
Table~\ref{tab:line_para}, these trends are also seen when we do the
bootstrapping analysis using 3$\sigma$ clipped weighted mean
statistics. However, the difference in the total \lya luminosity and
$\rm L^{r}_{\lambda}$ between two sub-samples are weaker when we use
3$\sigma$ clipped weighted mean stacked spectrum instead of median
stacked spectrum.  Similar trends are also seen if we consider
  the results for $CNR \ge 4$ cut-off [see
  Table~\ref{tab:line_para_withrelax_apdx} in the Appendix]. 
  Furthermore, we have tested the dependence of the strength of the
  \lya emission on $W_{1526}$ cut-off by decreasing it to
  $0.4$~\AA\ and $0.6$~\AA\, and find no strong dependence on this
  cutoff (see Table~\ref{tab:line_para_withrelax_apdx},
  Fig.~\ref{fig:profile_cnr5_ewcomp_appx}). \par

\emph{Results presented here are consistent with \lya emission being
  stronger among systems having large $W_{1526}$}. We also note that
the distribution of $N$(H~{\sc i}) and $z$ are similar (see the quoted
median values in Table~\ref{tab:line_lum_para}) between different
sub-samples discussed here purely based on $W_{1526}$. We discuss the
implications of these results in details in the
Section~\ref{lab:discussion}.

\begin{table*}
 \centering
 \begin{minipage}{150mm}
 {\scriptsize
 \caption{\small{\lya luminosity for the sub-sample with $W_{1526}
     \ge 0.4$~\AA\ and various $(r-i)$ color-cut.}}
 \label{tab:line_para_dust}
 \begin{tabular}{@{} r r c c r r r   c c c @{}}
 \hline 
  \multicolumn{1}{c}{Sample}  & \multicolumn{1}{c}{($r-i$)}  & \multicolumn{1}{c}{$\langle \Delta (r-i) \rangle$}  & \multicolumn{3}{c}{\lya luminosity }  &\multicolumn{2}{c}{velocity-offset}  &FWHM\\
                             &&&  \multicolumn{3}{c}{($\times 10^{40} \rm erg~s^{-1}$)}&&&\multicolumn{1}{l}{$\rm ( km\ s^{-1})$}\\
 
                              &&& DLA-bottom  & \multicolumn{1}{c}{L$_\lambda^b$}   &   \multicolumn{1}{c}{L$_\lambda^r$} &  \multicolumn{1}{c}{$\Delta v^r$} & \multicolumn{1}{c}{$\Delta v^b$} &     \\
      
\hline

lower 30\%  & $< 0.05$  &-0.06 & $22.89 \pm 7.21 $ & $ 6.02 \pm 4.79$ & $16.87 \pm 5.38$ &    $298\pm  30$& $-368\pm 15$& $196 \pm 70$\\ 
lower 40\%  & $< 0.08$  &-0.04 & $21.29 \pm 6.38 $ & $ 7.05 \pm 4.28$ & $14.23 \pm 4.74$ &    $336\pm  26$& $-363\pm 16$& $144 \pm 62$\\
lower 50\%  & $< 0.11$  &-0.03 & $14.00 \pm 5.82 $ & $ 2.71 \pm 3.92$ & $11.29 \pm 4.30$ &    $272\pm  46$& $-337\pm 26$& $214 \pm 98$\\
upper 50\%  & $> 0.11$  & 0.10 & $13.54 \pm 5.61 $ & $ 0.34 \pm 4.04$ & $13.20 \pm 3.89$ &    $      -   $&$      -   $& $      -       $\\
upper 40\%  & $> 0.13$  & 0.10 & $20.76 \pm 6.24 $ & $ 2.96 \pm 4.55$ & $17.80 \pm 4.28$ &    $      -   $&$      -   $& $      -       $\\
upper 30\%  & $> 0.16$  & 0.16 & $23.15 \pm 7.18 $ & $ 6.08 \pm 5.37$ & $17.06 \pm 4.77$ &    $      -   $&$      -   $& $      -       $\\

\hline                                                   
 \end{tabular} 
 }                                                   
 \end{minipage}
 \end{table*}

\begin{figure}
  \epsfig{figure=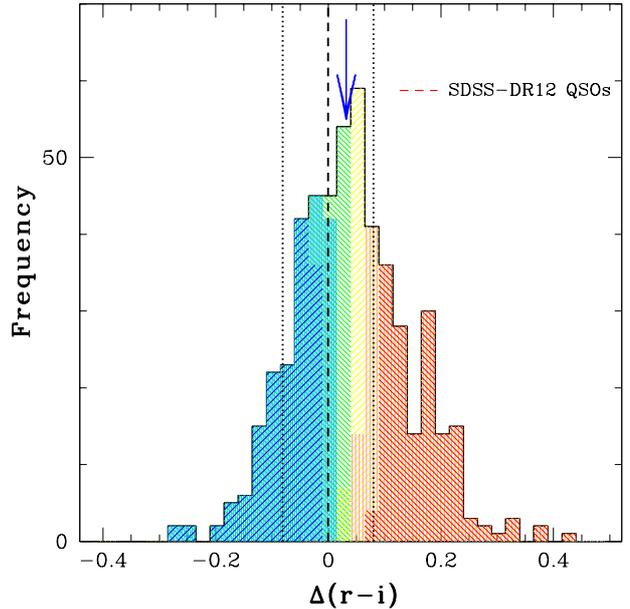,width=8.5cm,height=8.5cm}
  \caption{ 
    $\Delta$($r-i$) values of the background quasars for the
    sub-sample with $W_{1526} \ge 0.4$~\AA, with a median value is
    shown with a blue arrow. The $\Delta$($r-i$) distribution for
      the lower and upper 30, 40, 50\% of the sub-samples from ($r-i$)
      color distribution are shown with \emph{blue, cyan , green,
        yellow, orange and red} color, respectively. The dashed line
    represents the median $\Delta$($r-i$) color for the SDSS-DR12 QSOs
    with a similar redshift range  of the quasars used in the
      present work (i.e., $2.38 \le$ \zem $ \le 3.95$). The dotted lines
    show the one standard deviation away from that median.}
  \label{fig:dusthist}
    \end{figure} 

 \begin{figure*}
 \epsfig{figure=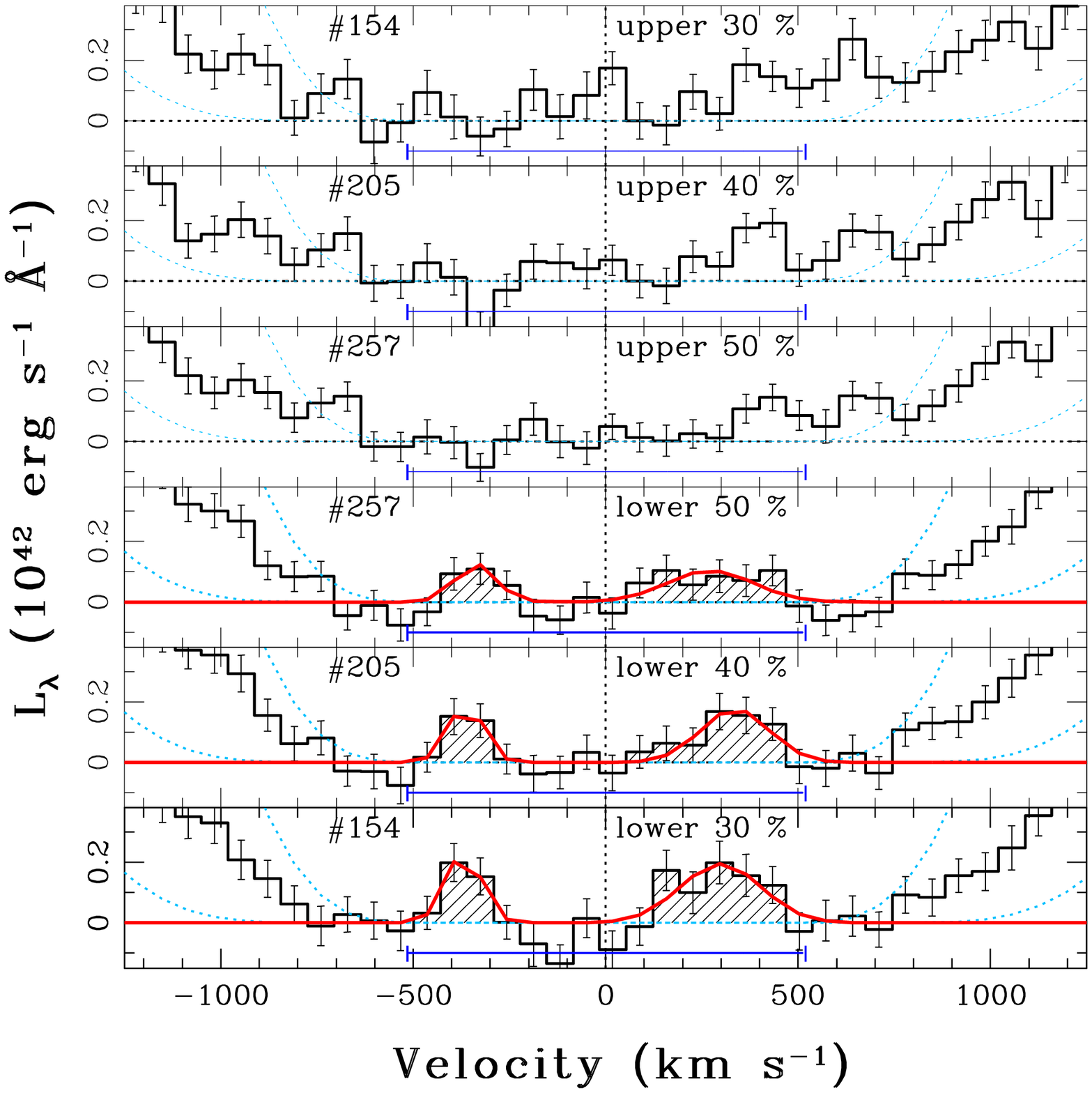,width=8.5cm,height=10.2cm}
 \epsfig{figure=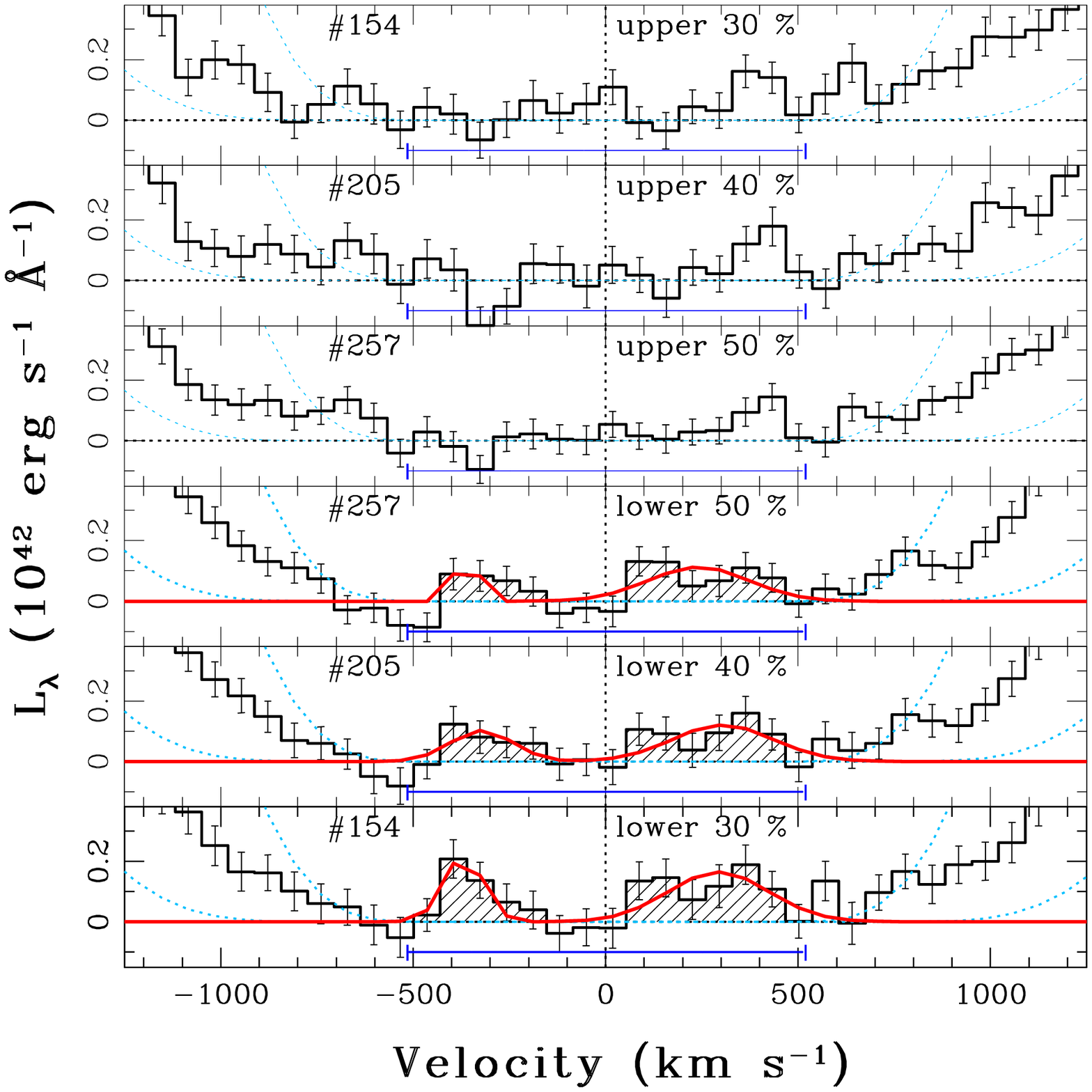,width=8.5cm,height=10.2cm}
 \caption{\emph{Left panel (bottom to top) :} The median stacked
   spectra for various sub-samples with $W_{1526} \ge 0.4$~\AA\ with
   color selection criteria of ($r-i$) $< 0.05, < 0.08, <0.11, > 0.11,  > 0.13$,
   and $>0.16$. The blue segment shows the DLA core with $\tau \ge 10$
   for log~\nhi\ $\rm = 21.0$. The dashed curves show the
   synthetic profiles for lower (i.e., log~\nhi $= \rm 21)$
   and median (i.e., log~\nhi $\rm = 21.23)$ column density of
   DLAs used to get the stacked spectrum. The solid red curve show the
   Gaussian fit to the non-zero flux seen in the red and blue part of
   the DLA core region. \emph{Right panel :} Same as left panel for
   3$\sigma$ clipped weighted mean stacks. See
   Fig.~\ref{fig:duststack_NHI21_cnr4_appx} in the Appendix for
   results when we consider systems with CNR $\ge 4$. In each panel
   the number of DLAs contributing to the plot are also indicated.}
 \label{fig:duststack}
   \end{figure*}

\subsection{Dependence of QSO colours:}
\label{subsec:color}

 The general population of DLAs tend to have very little dust so they
usually do not cause a strong reddening to the spectra of background
QSOs \citep[see][]{Vladilo2008A&A...478..701V,
  Frank2010MNRAS.406.2235F, Khare2012MNRAS.419.1028K,
  Fukugita2015ApJ...799..195F, Murphy2016MNRAS.455.1043M}. However,
due to the resonant nature of the \lya transition, presence of even a
small amount of dust can modify the emerging \lya line intensity and
profile shape. In addition, the spatial distribution of dust can also
change the profile of the emerging \lya emission. \par

In this section, we study the possible connection between the QSO
colour and the stacked \lya emission. For a given dust content in DLAs
the observed color of a QSO at a given redshift will also depend on
the absorption redshift. We consider two sub-samples based on ($r-i$)
colour of the QSO. \par

From Table~\ref{tab:line_lum_para}, we see that when we divide our
whole sample into two based on ($r-i$) colours there is no
statistically significant difference in the \lya luminosity between the two sub-samples even
though we see slightly higher value of the \lya luminosity (the
difference is only at 1.1$\sigma$ level) for the redder QSOs when we
consider the median stacked spectrum. From Table~\ref{tab:line_para},
we also notice that the red part shows a high luminosity of $\rm
L^r_{\lambda} = (9.8 \pm 3.5) \times 10^{40} erg\ s^{-1}$ (at $\sim
2.8 \sigma$) for the high ($r-i$) sample. Whereas, the low ($r-i$)
sample show a $\rm L^r_{\lambda} = (4.2 \pm 3.6) \times 10^{40}
erg\ s^{-1}$ with an insignificant (only at $\sim 1.2\sigma$)
luminosity difference between the two sub-samples. One needs to be
careful in interpreting these results as the difference in ($r-i$)
between the two sub-sample is not much and also there is a possible
correlation between ($r-i$) and W$_{1526}$. Just to explore the role
played by the possible presence of dust we restrict our future
discussions to systems with $W_{1526} \ge 0.4$~\AA. This is because we
already have a clear detection of \lya emission in this case. Thus
further dividing sub-sample based on ($r-i$) is more relevant for
probing the connection between reddening and \lya emission profile.
\par

 We compute the $\Delta(r-i)$ colours of QSOs having DLAs in our
 sample with W$_{1526}\ge0.4$~\AA\ by finding the deviation of the
 ($r-i$) colour of the QSO with respect to the median ($r-i$) colour
 of SDSS-DR12 QSOs at the same redshift. The $\Delta(r-i)$ is found to
 be strongly correlated with ($r-i$) with a $r_s$ of 0.88. In
 Fig.~\ref{fig:dusthist}, we show the histogram of $\Delta(r-i)$. The
 vertical dashed and dotted lines show the median $\Delta(r-i)$ and
 associated 1$\sigma$ spread for the general QSOs with the same
 redshift range of quasars used in the present work. The QSOs in the
 DLA sample are slightly more redder than the general population of
 QSOs. The presence of QSOs with negative $\Delta(r-i)$ indicates the
 possible intrinsic spread in the QSO continuum shape. We divide the
 W$_{1526}\ge0.4$~\AA\ sub-sample into 6 different groups by selecting
 the lower 30, 40, 50 and upper 50, 40 and 30\% DLA systems from the
 cumulative distribution of measured ($r-i$) color. Note that, the
 average $W_{1526}$ for the lower and upper 30\% sub-samples are found
 to be $0.83\pm0.13$\AA\ and $1.04\pm0.16$\AA, respectively. In
 Table~\ref{tab:line_para_dust}, we summarize our results. First six
 columns summarize the sample, range in ($r-i$) considered, mean
 $\Delta(r-i)$ of the group, the total \lya luminosity, L$_\lambda^b$
 and L$_\lambda^r$. More information can be obtained from the visual
 inspection of the profiles given in Fig.~\ref{fig:duststack}.

It is clear from Table~\ref{tab:line_para_dust} that when we consider
the stacked spectrum of lower 30 per cent of the quasars from ($r-i$)
color distribution, the integrated \lya luminosity over the full
profile is found to be (22.9$\pm$7.2)$\times 10^{40}$ erg s$^{-1}$.
Unlike all the other cases discussed till now, we do see a clear
double hump profile for the \lya emission in this case (see the bottom
panels of Fig.~\ref{fig:duststack}). Double hump feature for the lower
30 per cent of the quasars from ($r-i$) color distribution is also
clearly seen even in the 3$\sigma$ clipped mean stacked spectrum (see
right lower most panel in Fig.~\ref{fig:duststack}). It is also
consistent with the finding of \citet{Noterdaeme2014A&A...566A..24N},
where excess \lya emission is seen in extremely strong DLAs (i.e.,
log\nhi $\ge 21.7$) with lower $E(B-V)$ (i.e., $< 0.025$). Results of
double Gaussian fit to the observed \lya profile for the median
stacked spectra are also summarized in Table~\ref{tab:line_para_dust}
(see columns 7$-$9). From Table~\ref{tab:line_para_dust} we notice
that in the two low ($r-i$) bins within the measurement errors
L$_\lambda^r$ is twice that of L$_\lambda^b$ and the blue peak has a
slightly larger velocity shift with respect to the DLA redshift
compared to the red peak (see columns 7 and 8 in
Table~\ref{tab:line_para_dust}). Table~\ref{tab:line_para_dust} also
gives the deconvolved FWHM of the red peak obtained through Gaussian
fitting (column 9). In section \ref{sub:lyaprofile} we use these
values to compare DLAs with high redshift LBGs and \lya emitters
(LAEs). \par

It is clear from the bottom three panels in Fig.~\ref{fig:duststack}
and top three columns in Table~\ref{tab:line_para_dust} that as we
keep adding DLAs along the QSOs with high ($r-i$) colors the measured
\lya luminosity over the DLA core as well as L$_\lambda^b$ and
L$_\lambda^r$ decrease. In addition the significance of blue as well
as red peak decreases. This trend is in line with our expectation that
the presence of dust in DLAs will destroy the \lya photons. Based on
this trend, naively we would expect the double hump to disappear when
we consider sub-sample of DLAs towards QSOs having ($r-i$) in the
upper 50 per cent. While this is the case (see the second panel from
top in Fig.~\ref{fig:duststack}), we notice that the total integrated
\lya luminosity over the core region is similar to what we find for
the lower 50 per cent case. This is mainly because of the detection of
non-zero luminosity (detected at $\sim 3.4\sigma$ level) in the red
part. Indeed, when we consider only the upper 30 per cent of
sightlines based on ($r-i$) colours the \lya luminosity we measure is
close to what we measure for the lower 30 per cent case albeit without
a clear detection of the double hump.   \emph{While the
    disappearance of the double hump (or decreasing flux in the blue
    part) with increasing ($r-i$) is consistent with the effect of
    dust the lack of systematic change in the integrated \lya
    luminosity suggests that there are other effects at play while we
    change ($r-i$)}.

 It is to be remembered that when we consider the upper 30 per cent we
 probably have a mixed population of QSOs i.e., QSOs with bluer
 intrinsic colour reddened by the dust in the DLAs and QSOs that are
 intrinsically redder compared to the median QSO spectrum. Even if
 \lya is present in all these cases the profile shapes of the two
 populations are expected to be different that can lead to an
 incoherent addition and lack of a well defined shape for \lya
 emission. In addition, from figure 3, of
 \citet{Krogager2012MNRAS.424L...1K} one can see that the high H~{\sc
   i} column density systems may originate from smaller impact
 parameter when the metallicities are low. If this is the case with
 the low ($r-i$) sub-sample then the fiber filling factor will also
 have a role to play. We also notice a possible correlation between
 $\Delta(r-i)$ and $W_{1526}$. As seen in the previous section,
 systems with high $W_{1526}$ tend to have high \lya luminosity.
 Therefore, in the sub-samples based on $(r-i)$ it is not necessary
 that the only quantity that varies is the amount of reddening.
 Therefore interpretation of these results is not straight forward.

 \section{Absorption lines in the stacked spectrum}
\label{sec:absline}
  We also generated sets of median stacked spectra to study absorption
  lines, as described in section~\ref{lab:analysis}, for various
  sub-samples discussed in the previous section.  We detect
    transitions of the low ionization species (e.g. Fe{\sc~ii},
  Si{\sc~ii}, Zn{\sc~ii}, Cr{\sc~ii}, Mg{\sc~i}), high-ionization
  species (Si{\sc~iv}, C{\sc~iv}) and weak transition lines
  (Fe{\sc~ii}$\lambda \lambda$2249,2260) as found earlier in the
  stacked SDSS spectra by \citet{York2006MNRAS.367..945Y,
    Khare2012MNRAS.419.1028K} and
  \citet{Noterdaeme2014A&A...566A..24N}. In addition, we clearly
  detect Fe{\sc~ii}$\lambda1611$, Ni{\sc~ii}$\lambda1454$ and
  Ti{\sc~ii}$\lambda1910$ absorption which were marginally detected in
  the stacked spectrum of extreme-DLA systems studied by
  \citet{Noterdaeme2014A&A...566A..24N}. The rest equivalent width
  (\ew) measured for various metal lines in the stacked spectrum of
  our entire sample as well as different sub-samples, with a single
  Gaussian fit, are listed in Table~\ref{tab:line_EW}. Comparison of
  profiles of different absorption lines seen between different
  sub-samples are also shown in Fig.~\ref{fig:vel_plot_appx}.  We
    note that most lines are saturated even if the apparent optical
    depths of the strong absorption lines are small and appear
    unsaturated due to insufficient BOSS spectral resolution ($R \sim
    2000$) and smoothing resulting from redshift uncertainties when
    co-adding the spectra \citep{York2006MNRAS.367..945Y,
      Sardane2015MNRAS.452.3192S,Noterdaeme2014A&A...566A..24N}. \par

  \begin{figure*}
  \epsfig{figure=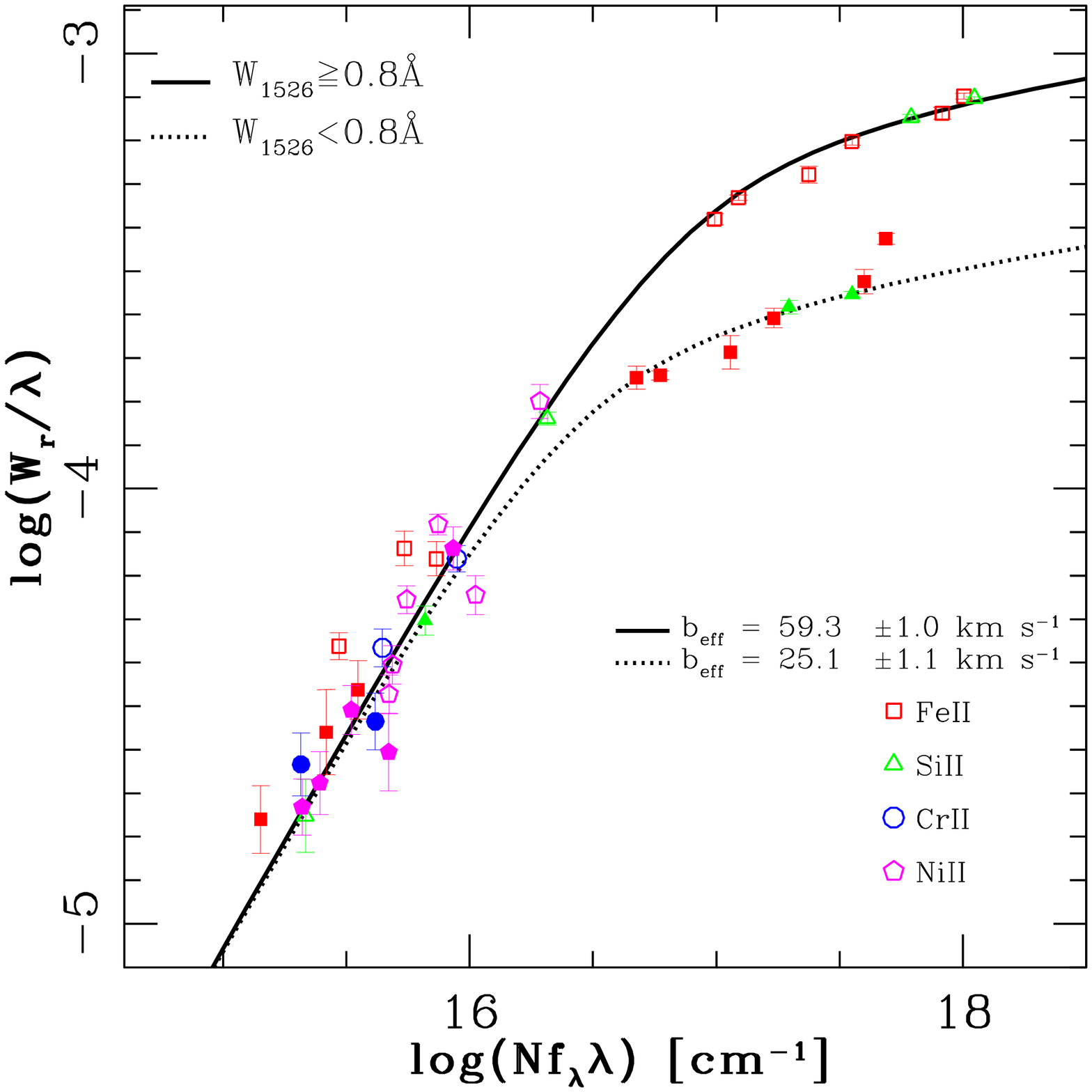,height=8.5cm,width=8.5cm}
  \epsfig{figure=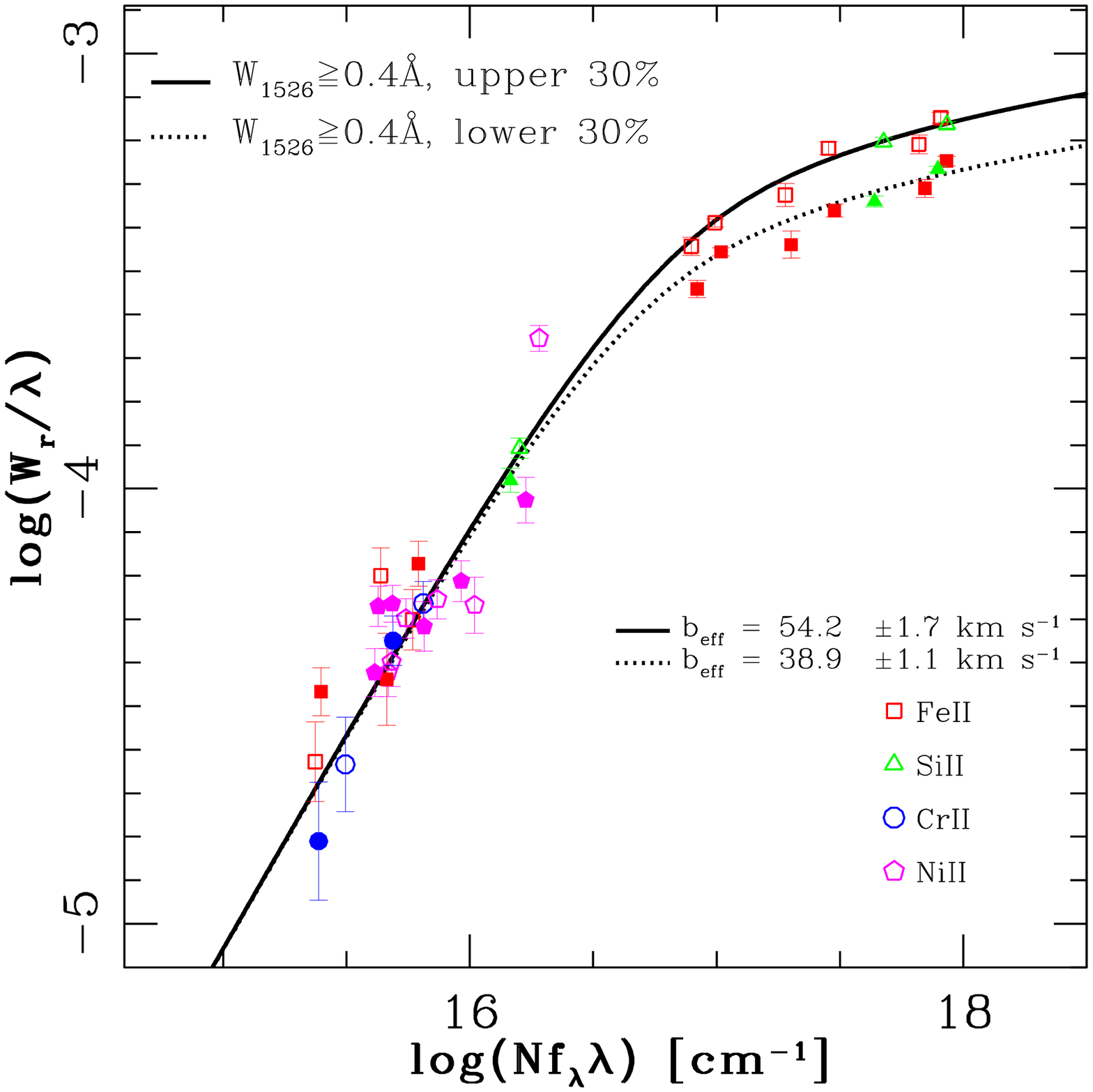,height=8.5cm,width=8.5cm}
   \caption{\emph{Left panel:} Single component curve of growth for
     absorption lines detected in the median stacked spectrum for the
     sub-samples with $W_{1526} \ge 0.8$~\AA\ (open squares and solid
     line) and $< 0.8$~\AA\ (filled squares and dotted line). The best
     fitted $b_{\rm eff}$ obtained are also quoted in the figure.
     \emph{Right panel:} Same for the stacked spectrum of $W_{1526} \ge
     0.4$~\AA\ and considering upper and lower 30\% sample from
     ($r-i$) color distribution.}
  \label{fig:cog}
   \end{figure*}

 \begin{figure}
  \epsfig{figure=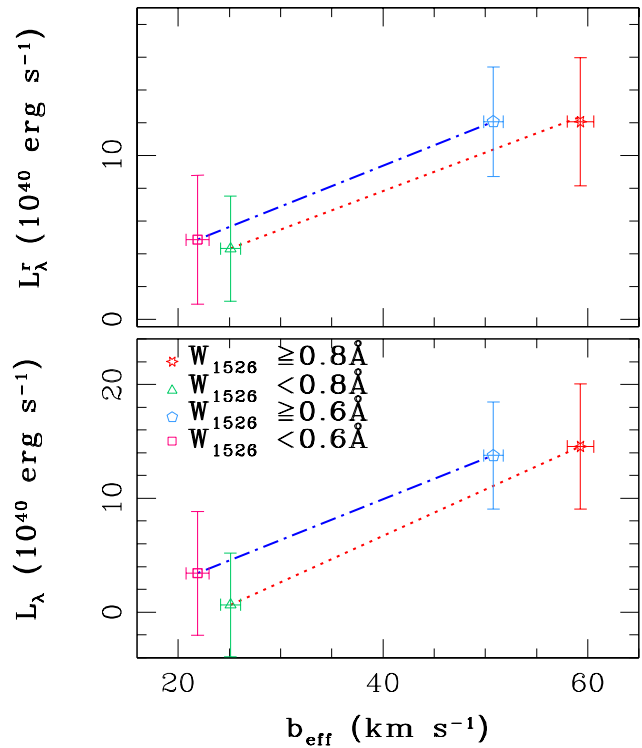,bbllx=18bp,bblly=162bp,bburx=328bp,bbury=532bp,clip=true,width=8.5cm,height=10.0cm}
 \caption{\emph{Lower panel:} The dependence of \lya luminosity
   (integrated over the entire DLA core region) on the Doppler
   parameter ($b_{\rm eff}$) for the sub-samples with $W_{1526} \ge$
   0.6, 0.8 and $< 0.6, 0.8$~\AA. \emph{Upper panel:} same for the
   luminosity seen in the red part of DLA bottom. Note that the
   significant detection (i.e., $> 3 \sigma$ level) is only for three
   cases with $W_{1526} \ge 0.6$ and 0.8~\AA.}
 \label{fig:si_vs_beff}
   \end{figure}

 \begin{figure*}
 \epsfig{figure=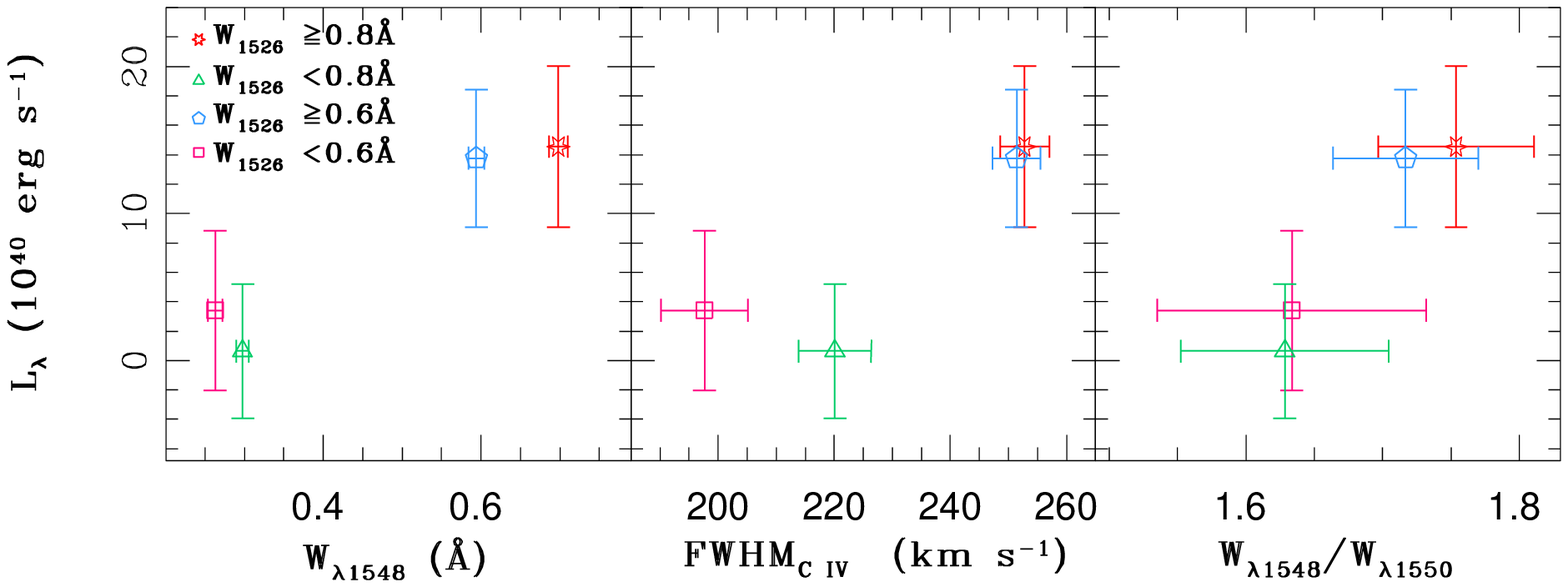,bbllx=20bp,bblly=155bp,bburx=572bp,bbury=374bp,clip=true,width=16.2cm,height=6.5cm}
 \caption{ The dependence of \lya luminosity (integrated over the
   entire DLA core region) on the equivalent width, FWHM of \civ
   absorption and the doublet ratio (i.e., $W_{1548}/W_{1550})$ for
   the sub-samples with $W_{1526} \rm \ge$  0.6, 0.8 and $<$ 0.6,
   0.8~\AA.}
 \label{fig:civ_vs_beff}
   \end{figure*}

  Next, using different absorption lines of \feii and \sii, that
  originate from transitions with a broad range of oscillator
  strengths detected in our stacked spectra, we compute the column
  density and the effective Doppler parameter ($b_{\rm eff}$) by
  constructing a single cloud curve growth (see left panel of
  Fig.~\ref{fig:cog} for the sub-samples based on W$_{1526}$). It is
  clear that $b_{\rm eff}$ is a function of $W_{1526}$.
  \citet{Jenkins1986ApJ...304..739J} has shown that a single cloud
  curve of growth (COG) gives nearly correct column density when using
  the \ew\ of several lines with the line distribution function not
  having markedly irregular characteristic, even when different lines
  have large variation in central optical depth and inter velocity
  dispersion. While spectral resolution of BOSS spectra will not allow
  us to measure $N$ and $b$ parameter accurately the purpose of using
  single cloud curve of growth is to find some tracers of column
  density and velocity field (i.e., $b_{\rm eff}$). \par

\begin{table*}                         
 \centering
 \begin{minipage}{150mm}
 {\small
 \caption{ \small{Parameters estimated using absorption stacks.}}
 \label{tab:absline_para}
 \begin{tabular}{@{} l c r c c c c c @{}}
 \hline 
 \multicolumn{1}{c}{sample}  & \multicolumn{1}{c}{$b_{\rm eff}$} & \multicolumn{1}{c}{$\Delta v_{90}$} & \multicolumn{1}{c}{$W_{1526}$} & \multicolumn{1}{c}{FWHM(\civ)} & \multicolumn{1}{c}{$DR${\textcolor{blue}{$^a$}}}  & \multicolumn{1}{c}{FWHM(\sii)}  \\
 \multicolumn{1}{c}{criteria}& \multicolumn{1}{c}{(km s$^{-1}$)} &\multicolumn{1}{c}{(km s$^{-1}$)}& \multicolumn{1}{c}{ (\AA)}& \multicolumn{1}{c}{(km s$^{-1}$)} & & \multicolumn{1}{c}{(km s$^{-1}$)}\\
      
\hline     
 $W_{1526} \ge 0.8$~\AA                 & $   59.3 \pm      1.3$  & 201  &  $1.21 \pm 0.01  $   & $  252.8  \pm  4.2$ &1.75 & $ 240.1 \pm 2.4 $  \\
 $W_{1526}  <  0.8$~\AA                 & $   25.1 \pm      0.8$  &  76  &  $0.43 \pm 0.01  $   & $  220.1  \pm  6.2$ &1.63 & $ 125.4 \pm 2.2 $  \\
 $W_{1526} \ge 0.6$~\AA                 & $   50.8 \pm      1.0$  & 164  &  $1.03 \pm0.01  $   & $  251.4  \pm  4.1$ &1.72 & $ 213.6 \pm 2.3 $  \\
 $W_{1526}  <  0.6$~\AA                 & $   21.9 \pm      0.6$  &  61  &  $0.37 \pm0.01  $   & $  197.7  \pm  7.5$ &1.63 & $ 114.0 \pm 2.6 $  \\
 $W_{1526} \ge 0.4$~\AA                 & $   45.3 \pm      1.0$  & 141  &  $0.85 \pm0.01  $   & $  250.3  \pm  4.0$ &1.74 & $ 192.7 \pm 2.2 $  \\
\nhi $\rm \ge 10^{21.23} ~cm^{-2} $     & $   34.4 \pm      1.0$  & 100  &  $ 0.75 \pm 0.01 $    & $  245.0  \pm  5.0$ &1.61 & $ 180.0 \pm 3.1 $  \\
\nhi $\rm <   10^{21.23} ~cm^{-2} $     & $   36.0 \pm      1.3$  & 106  &  $ 0.61 \pm 0.01  $   & $  237.7  \pm  6.4$ &1.71 & $ 167.4 \pm 3.6 $  \\
 \zabs $\ge 0.7$                       & $   37.2  \pm      1.8$ & 110  &  $ 0.65\pm 0.11  $   &  $  228.3  \pm  6.3$ &1.70 &  $ 171.3 \pm 3.6 $  \\
 \zabs $< 0.7$                         & $   37.8  \pm      1.0$ & 112  &  $ 0.69\pm 0.10  $   &  $  246.6  \pm  5.1$ &1.66 &  $ 175.0 \pm 3.1 $  \\
 ($r-i$) upper 30\%                    & $   54.2 \pm      1.7$  & 179  &  $ 1.05\pm 0.02  $   & $  240.5  \pm  6.5$ & 1.83 & $ 222.4 \pm 4.2 $  \\
 ($r-i$) lower 30\%                    & $   38.9 \pm      1.1$  & 116  &  $ 0.83\pm 0.01  $  & $  270.4  \pm  7.3$ & 1.68 & $ 184.6 \pm 3.5 $  \\
 \hline                                                   
 \end{tabular} 
 } 
\\
{ \textcolor{blue}{$^a$} {\civ doublet ratio, given as $DR = W$(C~{\sc
     iv}$_{1548}$)$/$$W$(C~{\sc iv}$_{1550}$).}}
                                                  
 \end{minipage}
 \end{table*}

  In Fig.~\ref{fig:si_vs_beff}, we show the integrated \lya luminosity
  in DLA core region (\emph{lower panel}) and in the red part of the
  DLA trough (\emph{upper panel}) versus $b_{\rm eff}$. In this figure, we
  mainly focus on the four sub-samples defined only based on
  $W_{1526}$. The $b_{\rm eff}$ value for various sub-samples are given in
  Table~\ref{tab:absline_para} (column 2). These values are much
  higher than what one measures in individual components seen in high
  resolution echelle spectra.  While comparing the $\Delta v_{90}$
    and $b_{\rm eff}$ measured from high resolution spectrum
    \citet{Noterdaeme2014A&A...566A..24N} have obtained a empirical
    relation of $\Delta v =2.2\ b_{\rm eff} + 0.02\ b_{\rm eff}^2$ which at
    larger $b_{\rm eff}$ departs from the linear theoretical relation in
    the Gaussian regime, $\Delta v = 2.33\ b_{\rm eff} $ (see their figure
    7). Therefore, $b_{\rm eff}$ we measure is an indication of overall
  velocity field and not related to the thermal or micro-turbulent
  motion. \par

   Interestingly we notice nearly the same $b_{\rm eff}$ (i.e.,
   between 35 and 38 \kms) for the sub-samples divided based on
   $N$(H~{\sc i}) and $z$ (not shown in this figure). Remember these
   sub-samples do not show significant difference in their \lya
   luminosities. Also, all of them tend to have less \lya luminosity
   compared to the high $W_{1526}$ sub-sample. Note,
   \citet{Noterdaeme2014A&A...566A..24N} have found a $b_{\rm eff} =
   40\ \rm km\ s^{-1}$ for their extremely strong DLAs with similar
   analysis. An increasing trend of \lya luminosity with $b_{\rm eff}$
   is clearly seen while considering the total luminosity in the DLA
   core as well as the ${\rm L_\lambda^r}$ (see
   Fig.~\ref{fig:si_vs_beff}). These trends are consistent with high
   $W_{1526}$ systems originating from gas with high velocity fields
   and also tend to show higher \lya luminosity. \par

   Similar dependences are also seen between [O{\sc~ii}] luminosity in
   the stacked spectrum and the equivalent width of Ca{\sc~ii} and
   \mgii absorption lines in the low-$z$ absorbers
   \citep{Wild2007MNRAS.374..292W,
     Noterdaeme2010MNRAS.403..906N,Menard2011MNRAS.417..801M}. This
   could be due to some physical connection between the star formation
   rate and velocity width (like for example large scale outflows as
   suggested by \citealt{Menard2011MNRAS.417..801M}) or an artifact of
   known correction between the line equivalent width and impact
   parameter coupled with fiber filling factor decreasing with the
   impact parameter and hence the rest equivalent width
   \citep{Lopez2012MNRAS.419.3553L}. Both these explanations will work
   for the \lya emission also. In addition, we also have the
   possibility that the \lya escape fraction being higher at higher
   \ew\ (i.e., velocity field). Note that, from the observed
   $W_{1526}$ and metallicity correlation the \lya emission will also
   depend on the metallicity. These trends are consistent with the
   suggestion of \citet{Moller2004A&A...422L..33M} based on DLA
   galaxies detected in their NICMOS sample. However, no such relation
   is seen with the $z$ and \nhi. \par

  Next we consider two sub-samples based on ($r-i$) colours of the
  background QSOs. As discussed in the previous section, the total
  \lya luminosity is nearly the same between the sub-samples
  consisting of the lower and upper 30 per cent sources. This was
  unexpected based on a simple conjecture that systems towards QSOs
  with high ($r-i$) colours and higher dust content should have less
  \lya luminosity. The stacked spectra show $b_{\rm eff}$ = 39
  \kms\ for the lower 30\% sub-sample and $b_{\rm eff}$ = 54 \kms\ for
  the upper 30\% sub-sample (see right panel in Fig.~\ref{fig:cog}).
  If one follows the idea of higher the $b_{\rm eff}$ higher will be
  the intrinsic \lya luminosity then the higher \lya luminosity in the
  upper 30\% can be understood as the consequence of systems having
  more intrinsic \lya luminosity.

  In Table~\ref{tab:absline_para} (see column 3), we present the
  velocity width ($\Delta v_{90}$) estimated from the relationship
  between $\Delta v_{90}$ and $b_{\rm eff}$ obtained by \citet[][see
    above]{Noterdaeme2014A&A...566A..24N}. There is a hint of excess
  \lya emission with the large values of $\Delta v_{90}$. It is clear
  that the inferred $\Delta v_{90}$ are more than 100 km s$^{-1}$ in
  cases where we have significant \lya detection. Interestingly, we
  note that in case of direct detection of \lya emission from DLA host
  galaxies the $\Delta v_{90}$ is found to be greater than $\sim$ 100
  $\rm km\ s^{-1}$
  \citep[see][]{Fynbo2010MNRAS.408.2128F,Noterdaeme2012A&A...540A..63N,Hartoog2015MNRAS.447.2738H,Srianand2016MNRAS.460..634S}.
        In a cosmological hydrodynamic simulations,
          \citet{Bird2015MNRAS.447.1834B} have shown that the velocity
          width closely tracks the virial velocity of the DLA host
          halo, and thus provides an analogue to the halo mass. In
          their models $\Delta v_{90} \sim 100$ \kms\ will corresponds
          to typically halo mass of $10^{10}-10^{11} \rm M_{\odot}$.
          In such halos higher gas accretion rate could lead to high
          SFR, hence higher intrinsic \lya and UV emission, albeit
          with a very small \lya escape
          fraction~\citep{Garel2015MNRAS.450.1279G}. Alternatively the
          observed low \lya luminosity be related to the diffuse \lya
          emission from the extended region of high mass galaxies. As
          we noted above the average luminosity is much less (i.e., 2
          to 2.5 per cent) than that of an L$_{\star}$ galaxy. We
          discuss these possibilities in the following section. \par

    We now investigate the dependence of \lya emission on the high
    ionization \civ line which probes a wider region of velocity space
    than the neutral lines \citep{Ledoux1998A&A...337...51L,
      Wolfe2000ApJ...545..603W, Fox2007A&A...473..791F}.
    \citet{Fox2007A&A...473..791F} have shown that high velocity \civ
    clouds are unbound to the central potential well and traces the
    outflows. In Fig~\ref{fig:civ_vs_beff}, we plot the \lya
    luminosity, integrated over the entire DLA bottom, as a function
    of \civ equivalent width, FWHM and the \civ doublet ratio (i.e.,
    $DR = W$(C~{\sc iv}$_{1548}$)$/$$W$(C~{\sc iv}$_{1550}$)) for the
    sub-samples based on $W_{1526}$. The measured line widths (FWHM),
    corrected for instrumental resolution, $DR$ and width of the
    \sii\ line are also summarized in Table~\ref{tab:absline_para}. As
    noted by \citet{Fox2007A&A...473..791F} \civ lines are wider than
    \sii\ lines. A trend of increasing \lya luminosity with increasing
    \ew\ and FWHM of \civ is clearly noticeable. In addition, the
    larger doublet ratio (i.e., optically thin gas) and higher
    equivalent width suggest that the winds from the galaxy are
    responsible for the absorption \citep{Bouche2006MNRAS.371..495B,
      Christensen2009A&A...505.1007C}. From
    Table~\ref{tab:absline_para}, it is clear that there is very
    little difference in \civ (FWHM) between the sub-samples defined
    based on ($r-i$) colours. The correlation between $L_{\lambda}$
    and $b_{\rm eff}$ is not present when we consider the two
    sub-samples based on ($r-i$) colours as discussed before. This
    once again reiterates the importance of dust and difficulties
    related to interpreting \lya profile in the stacked spectrum.

{\it In summary, the \lya emission from DLAs may be originating from
  systems with large velocity widths in both high and low ions. There
  are also indications that the metallicity and dust may have
  important role to play in deciding the total \lya luminosity and the
  profile shape. }

\section{Discussion}
\label{lab:discussion}

\begin{table*}
 \centering
 \begin{minipage}{150mm}
 {\small
 \caption{ \small{The star formation rate and surface brightness
     corresponding to the \lya luminosity seen in the red and blue
     part of DLA profile (median stack) in bootstrap analysis.}}
 \label{tab:line_para_sfr}
 \begin{tabular}{@{} l r  r r c c@{}}
 \hline 
 \multicolumn{1}{c}{Sample}    &\multicolumn{3}{c}{ SFR ($\rm M_{\odot} yr^{-1}$){\textcolor{blue}{$^a$}}} & \multicolumn{1}{c}{surface brightness{\textcolor{blue}{$^b$}}}\\
                              &&&& $(\times 10^{-19}\ \rm erg\ s^{-1} $\\                             
 
                              &  DLA-bottom &     blue-part          &   red-part &\multicolumn{1}{c}{$\rm cm^{-2} arcsec^{-2})$}\\
      
\hline

 Full with 90\% bootstarp                      & $ \le 0.55         $ &  $ \le 0.39 $   & $  \le 0.39     $ &   $ \le 3.93    $   \\  
 Full with 80\% bootstarp                      & $ \le 0.58         $ &  $ \le 0.41 $   & $  \le 0.42      $ &   $\le 4.17    $   \\  
 $W_{1526} \ge 0.4$ \AA                         & $ 0.71   \pm  0.21 $ &  $ \le 0.45 $   & $  0.65\pm  0.13 $ &   $6.66\pm 1.56$   \\  
 $W_{1526} \ge 0.8$ \AA                        & $  \le 0.87        $ &  $ \le 0.61 $   & $  0.64\pm  0.17 $ &   $6.51\pm 2.11$   \\ 
 $W_{1526}  <     0.8$ \AA                      & $ \le 0.85         $ &  $ \le 0.58 $   & $ \le 0.61       $ &   $\le 6.58$       \\  
 \nhi $\rm \ge 10^{21.23} ~cm^{-2}$            & $ \le 0.78         $ &  $ \le 0.55 $   & $ \le 0.56       $ &   $\le 5.51$       \\  
 $10^{21} \le$ ~\nhi $<    10^{21.23} ~cm^{-2}$ & $ \le 0.76         $ &  $ \le 0.53 $   & $ \le 0.55       $ &   $\le 5.53$       \\  
 \zabs  $\rm \ge 2.7$                          & $ \le 0.84         $ &  $ \le 0.58 $   & $ \le 0.60       $ &   $\le 4.79$       \\  
 \zabs  $\rm  <  2.7 $                         & $ \le 0.73         $ &  $ \le 0.51 $   & $ \le 0.52       $ &   $\le 6.43$       \\ 
 ($r-i$) $\ge 0.1$                             & $ \le 0.79         $ &  $ \le 0.56 $   & $ \le 0.55       $ &   $\le 4.98 $   \\
 ($r-i$) $< 0.1$                               & $ \le 0.78         $ &  $ \le 0.53 $   & $ \le 0.57       $ &   $\le 5.26 $      \\

\hline                                                   
 \end{tabular} 
 }{ \\ \textcolor{blue}{$^a$} a $3\sigma$ upper limit is given in case of
   non-detection.  Here, the errors are purely statistical and
     depends on the value of assumed $f_{esc}$ and IMF. See
     Section~\ref{subsec:sfr} for details of assumption involved in
     this estimate.}\\ { \textcolor{blue}{$^b$} Surface brightness
     corresponding to the \lya luminosity seen in the red part of DLA
     bottom.}
 \end{minipage}
 \end{table*}

In this section, we discuss implications of our \lya measurements for different
stacked spectra and draw some broad conclusions on the nature of high-$z$ DLAs.

\subsection{\lya emission line profile}
\label{sub:lyaprofile}
The \lya emission line profile is a powerful probe of the star
formation and associated feedback processes (i.e., infall and/or
outflows) in young galaxies. In an extremely opaque static medium \lya
escape through successive resonance scattering leading to a
double-humped profile, with the position of the peaks determined by
column density, temperature, and kinematics of the medium
\citep{Neufeld1990ApJ...350..216N,Neufeld1991ApJ...370L..85N,
  Verhamme2006A&A...460..397V, Hansen2006MNRAS.367..979H,
  Dijkstra2006ApJ...649...14D,Dijkstra2014PASA...31...40D}. In
addition, presence of bulk motions (e.g., outflow/inflow) and
scattering by dust in the H~{\sc i} gas can further modify the
emergent \lya profile. For instance, the scattering through an
outflowing (respectively inflowing) medium results in an overall
redshift (respectively blueshift) of the \lya spectral line with
enhanced red (respectively blue) peak and suppressed blue
(respectively red) peak \citep{Dijkstra2006ApJ...649...14D,
  Barnes2010MNRAS.403..870B}.

   We detect \lya emission predominantly in the red part of the
    \lya trough. This implies the presence of outflows in high-$z$
    DLAs. We detect flux in the red part at $2.8\sigma$ when we
    consider the full sample and with more than 3$\sigma$ level when
    we consider the sub-samples with $W_{1526}\ge~0.4$~\AA,
    0.6~\AA\ and 0.8~\AA\ (see Table~\ref{tab:line_lum_para} and
    Table~\ref{tab:line_para_withrelax_apdx}). In these cases, the
    average \lya luminosity is found to be consistent within the $\sim
    3\sigma$ upper limit of $\sim$ $21.8$ $\times 10^{40}$ erg
    s$^{-1}$ obtained by coadding 341 low column density (log~\nhi$
    \ge 20.62$) DLAs by \citet{Rahmani2010MNRAS.409L..59R}. However,
    less than the \lya luminosity of $(60.0\pm20.0)$ $\times 10^{40}$
    erg s$^{-1}$, seen in the composite spectrum of extremely strong
    DLAs (log~\nhi $\ge 21.7$) by
    \citet{Noterdaeme2014A&A...566A..24N}. We find the \lya luminosity
    being higher for systems with higher $W_{1526}$ which indicates
    that high metallicity DLAs could be associated with high star
    forming galaxies. Recently, the detection rate of DLA host
    galaxies is found to be higher towards high-metallicity DLAs
    \citep{Fynbo2010MNRAS.408.2128F,Fynbo2011MNRAS.413.2481F,
      Fynbo2013MNRAS.436..361F,Krogager2012MNRAS.424L...1K,Krogager2013MNRAS.433.3091K}.
    These galaxies show either no \lya emission or a suppressed blue
    peak.

  For the sub-samples with $W_{1526}\ge~0.4$ \AA, 0.6~\AA\ and 0.8~\AA\,
 with clear \lya detection, we model the red peak with a Gaussian
 profile and measure the centroid shift with respect to the redshift
 defined by the absorption lines detected in the quasar spectrum. As
 evident from Fig.~\ref{fig:stack_siew} the final profile shape of the
 \lya emission in the stacked spectrum depends on the statistics used.
       When we use median combined spectrum, we measure $\Delta
        v^r_{Ly \alpha}$ = $385 \pm 28$, $331 \pm 28$ and $337 \pm 32$
        \kms\ for the sub-samples with $W_{1526}\ge~0.4$ \AA,
        0.6~\AA\ and 0.8~\AA\ respectively. When we use the weighted
        mean spectrum, we get $\Delta v^r_{Ly\alpha}$ = $321\pm32$,
        $395\pm34$ and $295\pm56$ \kms\ for the three cases. From
      Table~2, we can see the mean $N$(H~{\sc i}) in all these cases
      are nearly the same i.e., log~$N$(H~{\sc i})$\sim$21.25. In a pure
      static medium, we expect $\Delta v^r_{Ly\alpha}$ in the range
      613$-$417 \kms\ if we assume H~{\sc i} gas temperature to be in
      the range, 10$^3$ to 10$^4$ K  \citep[][see their eq.
          21]{Dijkstra2014PASA...31...40D}.

Interestingly, the \lya emission line profiles associated with a
handful of confirmed DLA host galaxies are found to be predominantly
redshifted with typical velocity offsets ranging from $10-400$ \kms
\citep[][]{Moller1993A&A...270...43M,Moller2002ApJ...574...51M,Moller2004A&A...422L..33M,
  Fynbo2010MNRAS.408.2128F,Krogager2012MNRAS.424L...1K,
  Hartoog2015MNRAS.447.2738H,Srianand2016MNRAS.460..634S}. Thus, what
we measure in the stacked spectra are consistent with these individual
measurements. In addition, \citet{Noterdaeme2014A&A...566A..24N} have
shown that the \lya emission in the composite spectrum of extremely
strong DLAs also show a redshifted profile relative to the systemic
redshift. Till now only one DLA with \lya emission (\zabs = 2.207
towards the quasar SDSS~J113520.39$-$001053.56) shows a classical
double hump profile
\citep{Noterdaeme2012A&A...540A..63N,Kulkarni2012ApJ...749..176K}.
Even in this case detailed modelling of the \lya radiative transport
favors the emission being scattered from an outflowing gas
\citep{Noterdaeme2012A&A...540A..63N}.

In the case of high-$z$ LBGs and LAEs whenever systematic redshifts
can be determined using optical nebular emission lines the \lya
emission line profile is found to be asymmetric with the peak emission
being redshifted in the majority of the cases. The typical peak shift
measured is $\sim 200$ \kms\ in the case of LAEs
\citep{Shibuya2014ApJ...788...74S,Erb2014ApJ...795...33E,Song2014ApJ...791....3S,Trainor2015ApJ...809...89T}
and $\sim 400-600$ \kms\ in the case of LBGs
\citep[see][]{Steidel2003ApJ...592..728S,Steidel2010ApJ...717..289S,Kulas2012ApJ...745...33K}.
It has also been suggested that the shifts seen in the case of LAEs
and LBGs are statistically different. This could be attributed either
to the differences in $N$(H~{\sc i}) or to the H~{\sc i} covering
factor. Taken it at face value the $\Delta v^r_{Ly\alpha}$ we measure
for DLAs is higher than the mean value typically seen in the case of
LAEs but less than that of LBGs. \par

As discussed in the Section~\ref{subsec:color}, a clear double hump
profile is detected when we use DLAs detected towards bluer quasars.
The two peaks are separated $\sim 650 \rm\ km\ s^{-1}$, $\rm
L^{r}_{\lambda}/L^{b}_{\lambda} \sim 2$ and red part has an FWHM of
$\sim 200 \rm\ km\ s^{-1}$. About 30\% of LBGs
\citep{Kulas2012ApJ...745...33K} and 50\% of LAEs
\citep{Yamada2012ApJ...751...29Y,Hashimoto2015ApJ...812..157H} show
multiple peaks in their \lya emission with a typical peak separations
of $\sim 500 \rm\ km\ s^{-1}$ and $800 \rm\ km\ s^{-1}$ respectively,
for LAEs and LBGs \citep{Kulas2012ApJ...745...33K,
  Hashimoto2015ApJ...812..157H, Trainor2015ApJ...809...89T}. In
addition, the mean FWHM of the red peak of \lya line in LAEs and LBGs
are found to be $\sim$ 260 and 364 ~\kms, respectively
\citep{Trainor2015ApJ...809...89T}. The average property we measure
for the sub-samples with double hump \lya profile are intermediate
between LAEs and LBGs. However, it is interesting to understand the
connection between the appearance of a double hump feature and low
reddening of the quasars. A double hump feature is naturally produced
when \lya transport occur with nearly static \hi gas with very little
dust and weak bulk motions. Another possible explanation is that
systems towards blue quasars having large velocity infall. However, a
clear answer to our finding can only come from a detailed analysis of
individual systems with double hump detections in front of blue
quasars. \par

Some caution has to be exercised while comparing the DLA measurements
with those of LBGs and LAEs. \citet{Noterdaeme2010MNRAS.403..906N}
have detected nebular emission lines from $z\sim0.6$ Mg~{\sc ii}
systems in the SDSS fiber spectra of distant QSOs. From their figure
7, we can see that the relative velocity between the absorption
redshift and the redshift from the nebular line can be between $\pm
100$ \kms. If similar shifts exist between the line-of-sight where we
detect the DLA absorption and the star forming regions from where the
\lya emission originates then the intrinsic \lya emission profile will
be smeared by this additional random velocities. {\it While this
  effect prevents us from establishing a link between DLAs and either
  of LAEs or LBGs, the analysis presented here suggests that the
  average \lya profile we find follows the trends seen in the case of
  LBGs and LAEs. }

\subsection{Fluorescent \lya emission induced by UV background}

 We now explore the \lya emission  being induced by
metagalactic UV background impinging on the optically thick gas as a
possible mechanism for the \lya emission from DLAs. For this, we use
the recent computation of the background ionizing radiation by
\citet{Khaire2015MNRAS.451L..30K,Khaire2015ApJ...805...33K},
accounting for the contributions from both quasars and galaxies (with
typical escape fraction, $f_{esc}$, of 2 and 4 per cent for galaxies).
For the $f_{esc}$ of 0, 2 and 4 per cent, the integrated
unidirectional flux of the \hi ionizing photons (i.e., $\phi_0 =
\int_{\nu_{0}}^{\infty} \pi I_\nu d_\nu/h\nu$) at $z \sim 2.7$ are found to be $\sim$
9.1 $\times 10^{4}$, 1.3$\times 10^{5}$ and 1.7$\times 10^{5}$
photons\ $\rm s^{-1} cm^{-2}$, respectively
\citep[see][]{Shull2014ApJ...796...49S}. If we assume an arcsec$^2$
optically thick \hi gas cloud in photoionization equilibrium with this
background, it will produce a surface brightness of $\sim 3.1\times
10^{-20}, 4.6\times 10^{-20}$ and $6.2 \times 10^{-20} \rm erg\ s^{-1}
cm^{-2} arcsec^{-2}$. These values will be reduced if the clouds are
observed at some inclination angle.

Note that, in the presence of a local ionizing source the observed
surface brightness will increase and is related to surface brightness
induced by UV background as, $\rm SB = (1 + b) SB_{bgr}$. Here b is
the boost factor defined as the ratio of \hi\ photoionization rate
($\Gamma_{H I}$) from the local source to that of the UV background.
Table~\ref{tab:line_para_sfr} lists surface brightness corresponding
to the \lya luminosity measured in the red part or $3\sigma$ upper
limits. It is clear that, in cases where luminosity is measured at $>~
3\sigma$ level, the measured \lya surface brightness requires a boost
by a factor of more than 6 even if we consider meta galactic
UV-background computed with 4 per cent escape fraction for UV photons
from galaxies. In particular, for systems with $W_{1526} \ge \rm
0.8~\AA$ we need a boost in the \lya continuum flux by at least a
factor $\sim$10 from local sources of excitation probably from
\emph{in-situ} star formation or from star forming regions close to
the DLAs. In what follows, we discuss the implication of a local
radiation field in some detail.

 \begin{figure}
 \epsfig{figure=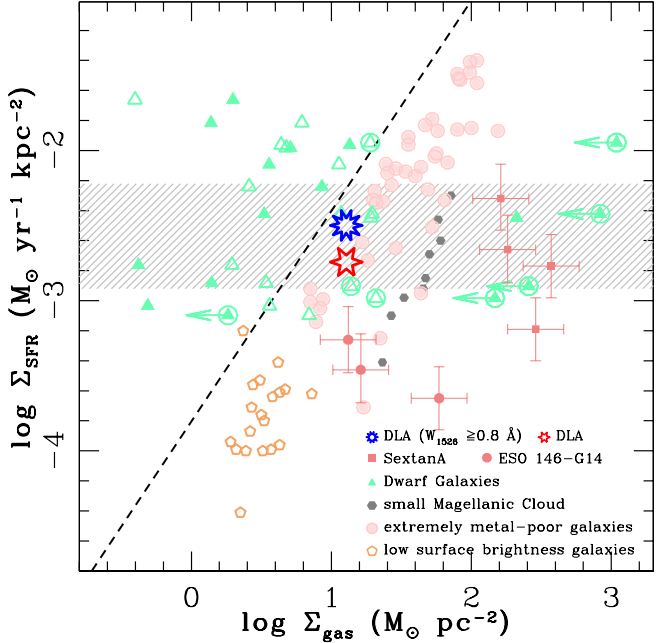,height=9cm,width=9cm}
 \caption{ The comparison of surface density of star formation and
   total (solid symbols) and HI(open symbols) gas surface density in
   DLAs with different classes of star forming galaxies. Orange points
   (\emph{open pentagon}) represent low surface brightness galaxies
   \citep{Wyder2009ApJ...696.1834W}, black points (\emph{small
     hexagon}) spatially resolved observations of Small Magellanic
   Cloud \citep{Bolatto2011ApJ...741...12B}, purple points
   (\emph{circle}) extremely metal poor galaxies
   \citep{Filho2016ApJ...820..109F}, green points (\emph{triangle})
   dwarf galaxies \citep{Cormier2014A&A...564A.121C}, brown points
   (\emph{square}) resolved observations of extremely metal poor
   galaxies SextanA and (\emph{circle}) ESO 146-G14
   \citep{Shi2014Natur.514..335S}. The dashed line corresponds to
   Kennicutt-Schmidt law \citep{Kennicutt2012ARA&A..50..531K}. The
   hatched region shows the surface brightness seen in the outskirts
   (i.e., $3-6$ kpc) of LBGs \citep{Rafelski2011ApJ...736...48R}. }
 \label{fig:ks_dwarf}
   \end{figure} 

\subsection{\emph{In-situ} star formation}
\label{subsec:sfr}

Assuming that the \lya photons mainly originate from H~{\sc ii}
regions around massive stars embedded in the DLAs and case
B-recombination \citep{Osterbrock2006agna.book.....O}, we relate the
\lya luminosity to the star formation rate SFR~$(\dot{M}_{\rm SF})$
as:
\begin{equation}
L_{\rm Ly\alpha} = 0.68 h\nu_\alpha (1-f_{esc}) N_\gamma \dot{M}_{\rm
  SF} .
\label{eq:sfr}
\end{equation}
Here, $h\nu_\alpha$=10.2~eV, is the energy of the \lya photons. 
At the redshift of interest in this work $f_{esc} \le 0.04$ based on
the $\Gamma_{H I}$ measurement using \lya absorption by the
intergalactic medium \citep{Khaire2016MNRAS.457.4051K}.
\citet{Vasei2016arXiv160302309V} have found a $f_{esc} \le 0.08$ using
deep HST imaging of a gravitationally lensed galaxy at $z \sim 2.38$.
In our calculations we use $f_{esc}=0.04$. The $N_\gamma$ represents
the number of ionizing photons released per baryons of star formation.
We use $N_\gamma = 9870$, corresponding to the average metallicity,
i.e. $Z/Z_{\odot} = -1.5$, of high redshift DLA absorbers and a
Salpeter initial mass function with $\alpha = 2.35$, given in
\citet[][references therein]{Rahmani2010MNRAS.409L..59R}. Note that
the observed \lya luminosity depends on the escape fraction
($f_{esc}^{Ly{\alpha}}$) of \lya photons and related to the emitted
\lya luminosity ($\rm L_{Ly{\alpha}}$) as $ L_{Ly{\alpha}}^{obs} =
f_{esc}^{Ly{\alpha}} L_{Ly{\alpha}}$. We use the
$f_{esc}^{Ly{\alpha}}$ to be 5 per cent, as is estimated for the
high-redshift ($z \sim 2.2$) star forming galaxies by
\citet{Hayes2010Natur.464..562H}. Based on the \lya to the H$\alpha$
luminosity ratio \citet{Trainor2015ApJ...809...89T} have recently
estimated the $f_{esc}^{Ly{\alpha}}$ to be 30 per cent for the LAE.
For our purpose, we use 5\% escape in our analysis.

The SFR corresponding to the \lya luminosity in the entire DLA bottom
as well as in the red and blue part for various sub-samples, measured
in our bootstrap analysis, are given in Table~\ref{tab:line_para_sfr}.
Note, if $f_{esc}^{Ly{\alpha}}$ is 30 per cent, as seen in the case of
LAEs, then the inferred SFR will be less than what is given in
Table~\ref{tab:line_para_sfr} by a factor 6. The average SFR in the
stacked spectrum of our full sample is found to be $\le$ 0.4 $\rm
M_{\odot} yr^{-1}$ in the red and blue part and also in the entire DLA
trough. However, the sub-samples with $W_{1526}\ge0.4$~\AA\ and
$W_{1526}\ge0.8$~\AA\ show a slightly higher SFR at $\sim$ 0.7~$\rm
M_{\odot} yr^{-1}$ for the total and the red part. In addition, for
the sub-samples based on \nhi\ and \zabs\ the $3\sigma$ upper limit on
SFR is found to be $\sim$ 0.6~$\rm M_{\odot} yr^{-1}$ (see also
Table~\ref{tab:line_para_sfr}). \par

 Next, we explore the Kennicutt-Schmidt (KS) relation between the gas
 surface mass density and the star formation rate per unit surface
 area in DLA galaxies. Assuming the typical size of the DLA galaxies
 to be $\rm R=8~kpc$ (the fiber filling factor of 1) we compute the
 average SFR per unit area to be $\rm log \Sigma_{SFR} = -2.7$ and
 $-2.5\ \rm M_{\odot} yr^{-1} kpc^{-2}$ in the red part of DLA bottom
 for our entire sample and the sub-sample with $W_{1526}\ge0.8$~\AA,
 respectively. As the fiber filling factor may be smaller than one,
 the measured $\rm \Sigma_{SFR}$ values should be a lower limits. For
 the mean log~\nhi$=21.23$, probed in this work, the expected average
 SFR is about $0.01\ \rm M_{\odot} yr^{-1} kpc^{-2}$, if one considers
 the KS relation for low redshift star forming galaxies
 \citep{Kennicutt1998ARA&A..36..189K,Kennicutt1998ApJ...498..541K}.
 The surface SFR densities in DLA galaxies are a factor of $\sim$ 6
 lower than what one would predict from the local star formation law
 \citep[see
   also,][]{Chelouche2010ApJ...722.1821C,Rahmani2010MNRAS.409L..59R,Rafelski2011ApJ...736...48R}.
 \par

In Fig.~\ref{fig:ks_dwarf}, we compare the gas surface density ($\rm
\Sigma_{gas}$) with the surface SFR ($\rm \Sigma_{SFR}$) in different
kinds of galaxies with what we measure for DLAs. It is clear from the
figure that the DLAs occupy the region that is generally populated by
low metallicity dwarf galaxies and extremely metal poor (i.e. $Z <
0.1Z_{\odot}$) galaxies
\citep{Wyder2009ApJ...696.1834W,Bolatto2011ApJ...741...12B,
  Shi2014Natur.514..335S,Cormier2014A&A...564A.121C,Filho2016ApJ...820..109F}.
In this figure we also compare the surface brightness seen in the
outer region of LBGs in the stacked images by
\citet{Rafelski2011ApJ...736...48R}. Our measurements and the limits
are consistent with this range. It is interesting to note that in the
case of DLAs despite log~$N$(H~{\sc i}) $\ge$ 21 we do not detect H$_2$
molecules in our stacked spectrum \citep[see][for H$_2$-H~{\sc i}
  transition in DLAs using individual H$_2$
  measurements]{Noterdaeme2015A&A...578L...5N} so the surface mass
density is mainly atomic. However, even in the case of dwarf galaxies
a considerable contribution to the measured surface mass density comes
from molecular gas. While H~{\sc i} emission is widespread in the
dwarf galaxies, regions with log~$N$(H~{\sc i})$\ge$21 project smaller
cross-section and usually localized to the star forming regions even
if there is no one to one correspondence between SFR and $N$(H~{\sc
  i})
\citep[]{Begum2006MNRAS.365.1220B,Roychowdhury2014MNRAS.445.1392R}.
Therefore, in this scenario the continuum and \lya emission associated
with the H~{\sc i} gas can be well within the SDSS fiber. \par

The low star formation efficiency in low metallicity dwarf galaxies
and extremely metal poor galaxies is mainly attributed to the (i) low
cooling rate in metal poor environments
\citep{Filho2013A&A...558A..18F, Filho2016ApJ...820..109F}; (ii)
galactic winds which can suppress the SFR by throwing more than 80\%
of gas back to the circum-galactic medium
\citep{Almeida2014A&ARv..22...71S} and/or (iii) related to the star
formation in compact regions which are fed by the accretion of metal
poor \hi gas by cold cosmological accretion
\citep{Ekta2010MNRAS.406.1238E, Almeida2014A&ARv..22...71S}. The same
may be the reason for low SFR seen in DLA galaxies as well, as the
DLAs are also found to have a low average metallicity, i.e., $\sim
1/10th$ of solar, at $z \sim 2$
\citep{Rafelski2012ApJ...755...89R,Noterdaeme2014A&A...566A..24N} and
signatures of outflows as an off-centered \lya emission in our stack.

 The low surface brightness galaxies or the low surface brightness
  outer regions of luminous galaxies contributing to the high-$z$ DLA
  population is a good possibility, even though at low-$z$ such
  galaxies do not contribute appreciably to the high $N$(H~{\sc i})
  absorbers studied here \citep{Patra2013MNRAS.429.1596P}. While we
  expect the number density of low mass galaxies to be much higher at
  higher $z$, \lya forest - DLA cross-correlation analysis suggests
  that most of the high-$z$ DLAs may originate from a biased regions
  with halos of 10$^{12}$ M$_\odot$
  \citep{Font-Ribera2012JCAP...11..059F}. \par

Next, we consider the possibility of DLAs originating from passive
H~{\sc i} gas at the external regions of massive star forming
galaxies. In this scenario, the \lya emission will be detected either
when the galaxy light falls inside the fiber (i.e., impact parameter
less than 8 kpc) or when the Lyman continuum photons from the host
galaxy inducing the \lya fluorescence emission from the extended gas
(i.e., the so called \lya halo). \par

 \begin{figure}
 \epsfig{figure=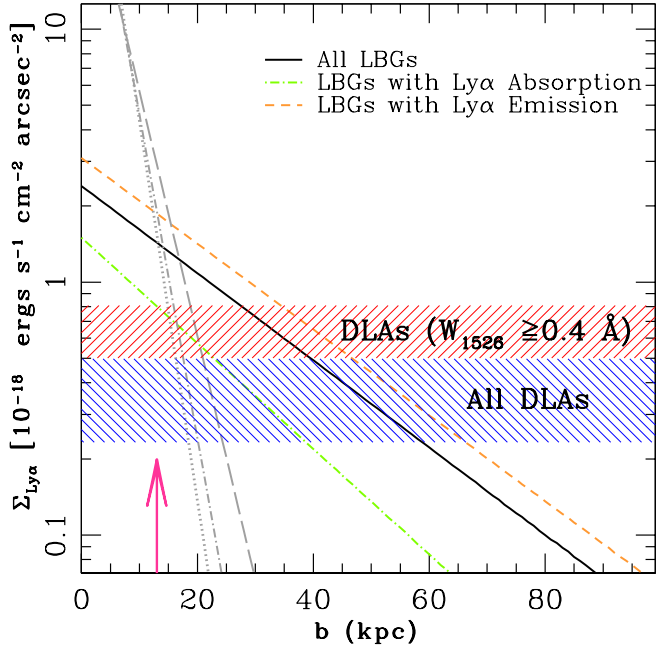,height=9.0cm,width=9.0cm,angle=0}
 \caption{The average surface brightness profile of the \lya emission
   in a stacked images of 92 LBGs (solid black) and in a sub-sample
   with LBGs seen in \lya emission (orange dashed) and in \lya
   absorption (green dot-dashed) by
   \citet{Steidel2011ApJ...736..160S}. The gray lines represent the
   average surface brightness profile of continuum light (at 1220~\AA)
   for all LBGs (dashed dotted) as well as the sub-samples with only
   \lya emission (dotted) and absorption (dashed). The hatched region
   show the surface brightness seen in our DLA stacks for entire
   sample (blue) and for a sub-sample (red) with $W_{1526} \ge
   0.4$~\AA, respectively. The arrow represent the expected
   $R_{\star}$ for unit covering factor of \hi gas around galaxies
   from discussion in Section~\ref{subsec:sfr_in_gal}.}
 \label{fig:sb_steidel}
   \end{figure}

\subsection{Star forming galaxies at small impact parameter}
\label{subsec:sfr_in_gal}

       The number of DLAs at $z\sim 2.7$ having log~\nhi $\ge
        21.0$ per unit absorption distance is estimated to be
        $dN/d\mathcal{X} \approx 0.028$, by integrating over the
        column density distribution function $f_{H I}(N, \mathcal{X})$
        from \citet{Noterdaeme2009A&A...505.1087N}. This can be
          used to derive the co-moving incidence of star forming
          galaxies at high redshift (LBGs) as:
\begin{equation}
d{\ensuremath{\mathcal{N}}}_{LBGs}(L>L_{min})/dl_c = (1+z)^2 f
\int_{L_{min}}^{\infty} \pi R(L)^2 \phi(L) dL.
\label{eq:lbg_lf}
\end{equation}
   Here, $R(L)$ represents size (or impact factor) of the \hi gas
   below which the required \nhi\ will be produced in a galaxy with
   luminosity $L$. We use a power-law relation, i.e., $R(L) \propto
   L^{0.4}$, obtained from \hi size-optical luminosity relation from
   field galaxies in ALFALFA survey which is determined based on the
   galaxy R-band luminosity \citep{Toribio2011ApJ...732...93T}. The
   $f$ factor is the gas filling factor which is considered to be one. We use
   the UV luminosity function of LBGs by
   \citet{Reddy2009ApJ...692..778R} at redshifts $1.9 \le z \le 2.7$,
   with best fit Schechter function parameter of $\rm \phi^{\star} =
   2.75 \times 10^{-3} Mpc^{-3}, M^{\star}_{AB}(1700 \AA) = -20.70$
   and $ ~\alpha= -1.73$. Using the above scaling relation, $R(L) =
   R_{\star}(L/L_{\star})^{0.4}$, the equation(\ref{eq:lbg_lf})
   becomes:
\begin{equation}
 \scriptsize{
\frac{d{\ensuremath{\mathcal{N}}}_{LBGs}}{dl_c}  = \pi R_{\star}^2 (1+z)^2 f \int_{L_{min}}^{\infty}   \left(\frac{\phi^{\star}}{L^{\star}}\right) \left(\frac{L}{L^{\star}}\right)^{(\alpha+0.8)} exp\left(-\frac{L}{L^{\star}}\right)  dL } 
\label{eq:lbg_lf_rlscale}
\end{equation}
 Owing to the large magnifications by a lensing cluster
 \citet{Alavi2014ApJ...780..143A} have probe the LBGs down to very
 faint magnitudes, $\rm M_{1500} < - 13~mag$, about 100 times fainter
 than previous studies at the same redshift
 \citep{Reddy2009ApJ...692..778R}. They have shown that the UV
 luminosity function of LBGs shows no turnover up to $\rm M_{1500} < -
 13~mag$ and have a faint end slope, $\alpha =1.74$, similar to
 \citet{Reddy2009ApJ...692..778R}. We compute the impact factor, $R
 \sim 13\ \rm kpc$, below which required \nhi\ will be produced by an
 $L^{\star}$ galaxy by integrating the above equation down to $L_{\rm
   min} = 0.001 L^{\star}$. It implies a high probability for the
 $L^{\star}$ galaxy to lie outside the small region of 8~kpc probed by
 the BOSS fiber. Alternatively, taking the maximum value of $R$ to be
 $\sim$8~kpc (corresponding to the fiber size) in above scaling
 relation $R = R_{\star} ( L/L^{\star}) ^\alpha$, we find the maximum
 luminosity of a galaxy which will come inside the fiber to be $0.3
 L^{\star}$. In other words, the majority of galaxies contributing
 directly to the average luminosity seen in stacked spectrum will be
 below $\sim$ 0.3$L^{\star}$. \par

The average luminosity of LBGs with
 luminosity ranging from $0.001-0.3 L^{\star}$ is given as:
 \begin{equation}
\rm \langle L_{LBG} \rangle = \frac {\int_{L_{min}}^{L_{max}} L
  \phi(L) dL} {\int_{L_{min}}^{\infty}  \phi(L) dL} .
\label{eq:lavg}
\end{equation}
  We find the average UV luminosity of LBGs to be $1.34\times 10^{42}
  \rm erg\ s^{-1}$, corresponds to a SFR of $\rm
  0.12~M_{\odot}~yr^{-1}$. Interestingly, this is comparable to the
  average SFR of $\rm 0.09 < SFR < 0.27\ M_{\odot}~yr^{-1}$ ($2\sigma$
  limit) seen in direct imaging of DLA host galaxies within 2 and 12
  kpc aperture sizes by \citet{Fumagalli2015MNRAS.446.3178F}, albeit
  for DLAs with lower \nhi\ than what we consider here. Note that if
  we assume the gas filling factor to be less than 1, then $R_{\star}$
  will be $>$ 13~kpc and $L_{max} < 0.3~L_{\star}$ and the average UV
  luminosity will be even lower than the above quoted value. The
  expected SFR within 12~kpc aperture, using the above formalism, is
  found to be $0.14~\rm M_{\odot}~yr^{-1}$. Therefore, allowing for
  the impact parameter for a given $N$(H~{\sc i}) to scale with
  optical luminosity (as seen in the local universe) for high-$z$ DLAs
  gives the mean SFR in the correct ballpark value. Converting the
  average UV luminosity estimates into \lya luminosity is not straight
  forward as the fraction of \lya emitting galaxies may be a function
  of optical luminosity. However, one can get a simple estimate by
  substituting our constraint on SFR in eq~\ref{eq:sfr}. If we use
  $f_{esc} = 0.04$ then in order to explain the \lya luminosity we
  find for high $W_{1526}$ sub-samples we need $f_{esc}^{Ly{\alpha}} =
  0.22$. This is in between the values found for LBGs (i.e, 5 per
  cent) and LAEs (i.e., 30 per cent). \par

In the above discussion, we found most of the \lya emission from bright
galaxies associated with a DLA will be outside the SDSS fiber.
However, even in this case \lya fluorescence from the large extended
regions can still contribute to the \lya emission inside the fiber.
Extended diffuse \lya emitting regions (up to 80 kpc radius) are
detected around high redshift LAEs and LBGs in recent observations by
stacking analysis \citep{Hayashino2004AJ....128.2073H,
  Steidel2011ApJ...736..160S,
  Matsuda2012MNRAS.425..878M,Feldmeier2013ApJ...776...75F,
  Momose2014MNRAS.442..110M}. The presence of these diffuse \lya
emitting halos (LAHs) are also predicted by numerical simulations
\citep{Laursen2007ApJ...657L..69L, Zheng2011ApJ...739...62Z,
  Dijkstra2012MNRAS.424.1672D,
  Verhamme2012A&A...546A.111V,Jeeson-Daniel2012MNRAS.424.2193J}. \par

\citet{Momose2016MNRAS.457.2318M} have found prominent LAHs around
LAEs with faint \lya luminosities, bright UV luminosities, and small
\lya equivalent width.  The extent of the LAHs is found to be
  dependent on the environment, where LAHs in low-density environment
  are found to be smaller than the higher density environments
  having a typical scale length of $\sim$ 9.1 and 20.4 kpc,
  respectively \citep{Matsuda2012MNRAS.425..878M,
    Momose2014MNRAS.442..110M}. In addition, in continuum-subtracted
\lya emission images of LBGs \citet{Steidel2011ApJ...736..160S} have
found similar diffuse \lya halos for all the galaxies, irrespective of
whether the \lya is seen in emission or absorption, up to radii of
$\sim$80 kpc. In Fig.~\ref{fig:sb_steidel}, we compare the average
\lya surface brightness profile found by
\citet{Steidel2011ApJ...736..160S} around the LBGs with our
measurements. Our discussions in the previous section suggests an
L$_{\star}$  galaxy will have a typical impact parameter of $\sim$ 13~kpc. It
is clear from the figure that surface brightness contributions at
these impact parameters will be much higher than the average we find
for our samples. Inclusion of this contribution will increase the \lya
luminosity inside the fiber from stars that do not contribute to the
continuum stacked image. Therefore, the $f_{esc}^{Ly{\alpha}}$ can be
smaller than 0.22 we infer to match our \lya luminosities with the SFR
constrains from the stacking experiment.

\section{Summary}

In order to probe the star formation in neutral gas clouds at high
redshift, we perform a stacking analysis of 704 DLAs with a large \hi
column density (log~\nhi$\rm \ge 21$) and a median redshift of
$\sim$2.7. We generate both emission and absorption line stacked
spectrum for various sub-samples based on \nhi, \zabs, $W_{1526}$ and
$(r-i)$ colours of QSOs which led us to the following conclusions :

(1) For the full sample, we measure the \lya luminosity of
$(5.2\pm3.3)\rm \times 10^{40} erg~s^{-1}$ when we integrate the
luminosity over the full core regions of the DLA in the median
spectrum. Similar values are obtained when we considered the stacked
spectra obtained using weighted mean and 3$\sigma$ clipped weighted
mean. This luminosity is $\le 0.1 $ per cent of the \lya luminosity of
$L_{\star}$ galaxies at these redshifts.  In the bootstrap
  analysis, the measured luminosities in the red and blue part of the
  core regions of the DLA troughs in the median stacked spectrum are
  $(-1.1\pm2.4)\rm \times 10^{40} erg~s^{-1}$ and $(6.9\pm2.5)\rm
  \times 10^{40} erg~s^{-1}$ respectively, with a $\sim$2.8 $\sigma$
  excess \lya emission in the red part of the DLA trough.

(2) In the sub-samples based on $W_{1526}$ we find the \lya
  luminosities of $(13.5\pm4.1)\rm \times 10^{40} erg~s^{-1}$ and
  $(14.6\pm5.5)\rm \times 10^{40} erg~s^{-1}$ when we integrate the
  luminosity over the core of \lya trough in the median spectrum for
  $W_{1526}\ge$ 0.4 and 0.8 \AA\ respectively. These luminosities are
  mainly contributed from the red part with respective luminosities of
  $L_\lambda^r = (12.3\pm2.9)\rm \times 10^{40} erg~s^{-1}$ and
  $(12.1\pm3.9)\rm \times 10^{40} erg~s^{-1}$. Blue parts have
  luminosities consistent with zero. For systems with Si~{\sc ii}
  detections having $W_{1526}<0.8$~\AA\ we find the \lya luminosity to
  be $(0.6\pm4.5)\rm \times 10^{40} erg~s^{-1}$. Even in this case
  most of the observed luminosity comes from the red part albeit with
  a significance of 1.3$\sigma$ level. As the median $N$(H~{\sc i}) in
  all these sub-samples are the same, the difference in the \lya
  luminosity are not likely related to the differences in the \hi
  optical depth. \citet{Prochaska2008ApJ...672...59P} have found a
  strong correlation between $W_{1526}$ and metallicity using echelle
  spectroscopic data. This correlation has been interpreted as a
  mass-metallicity relation as high $W_{1526}$ absorption tend to
  trace low optical depth clouds in the halo or outflowing gas. Thus
  the enhanced \lya luminosity in the high $W_{1526}$ could be an
  indication of high star formation in high metallicity systems
  together with easier escape of \lya photos enabled by the large
  outflowing gas. We measure large $b_{\rm eff}$ for systems with high
  $W_{1526}$ which supports this hypothesis.

(3)  The sub-samples based on $N$(H~{\sc i}) do not show any
detectable difference in the measured \lya luminosities either in the
full core region or in the red part alone. The same is the case when
we divided the sample into two redshift bins. We note that all the
four sub-sample have nearly similar mean $W_{1526}$ (see
Table~\ref{tab:line_EW}).

(4) The sub-samples based of $(r-i)$ colours and $W_{1526} >
0.4$~\AA\ show a double hump profile for the low $(r-i)$ sub-sample.
The double hump disappears when we add more and more red QSOs
sightlines to this sample. However, total \lya luminosity does not
show any monotonous trend with $(r-i)$.  In addition, the $\rm
  L^r_{\lambda}/L^b_{\lambda}$, FWHM and peak separation are
  intermediate between what is seen in LBGs and LAEs. However to
  uniquely identify the reason behind the appearance of double humped
  \lya line toward blue QSOs we need detailed analysis of such
  individual systems using radiative transfer models. Establishing
any trend between the dust indicators and \lya profile will help us to
discriminate between different models, such as (i) static medium
\citep{Neufeld1990ApJ...350..216N,
  Zheng2002ApJ...578...33Z,Dijkstra2006ApJ...649...14D}, (ii)
expanding/inflowing shell
\citep{Ahn2003MNRAS.340..863A,Verhamme2006A&A...460..397V,Schaerer2011A&A...531A..12S}
or (iii) multiphase media \citep{Gronke2016ApJ...826...14G}, of the
\lya radiative transport in DLAs. As these models predict different
shape as a function of dust content.

(5)  In our full sample as well as in all the sub-samples, we
detected \lya emission predominantly in the red part of the \lya
trough albeit with varied significance. The measured shift in the peak
location of \lya emission with respect to the absorption redshift is
in the range 300-400 \kms. The redshifted profiles are typical of what
is seen in the case of LBGs and LAEs at the same redshifts. In these
cases the observed redshifted \lya profiles are considered as a
signature of predominantly outflowing gas in these galaxies. Thus the
\lya profile we measure in the case of high-$z$ DLAs are consistent
with the presence of outflowing gas. The measured shift in DLAs are
higher than what is typically seen in LAEs but less than those of
LBGs. However, as our reference redshift is the absorbing gas along
the QSO sightline and not the non-resonant nebular emission from the
galaxy itself, we need to exercise caution in interpreting this
result.

(6) Using the updated metagalactic UV background radiation contributed
by QSOs and galaxies, we find the expected \lya fluorescence is 4 to
10 times less than what we measure. This means most of the DLAs are
not passive clouds in ionization equilibrium with the metagalactic UV
background. Local excess ionizing radiation either from a nearby star
forming region or from \emph{in-situ} star formation is needed to
produce the observed \lya luminosities. We discuss different scenarios
such as \emph{in-situ} star formation in low luminosity galaxies or
scattered \lya emission from an extended \lya halos around high
luminosity galaxies as possible alternatively to explain the observed
\lya luminosity. If the bias estimated for the DLA population based on
clustering analysis as well as our mass estimation from the stacked
absorption lines are true then the second alternative will be favored.

\section*{Acknowledgments}
 We are grateful to Vikram Khaire for providing the metagalactic UV
 background. RS, PN, and PPJ acknowledge the support from Indo-French
 Centre for the Promotion of Advance Research (IFCPAR) under project
 number 5504$-$2.\par

 Funding for the SDSS and SDSS-II has been provided by the Alfred P.
 Sloan Foundation, the Participating Institutions, the National
 Science Foundation, the U.S. Department of Energy, the National
 Aeronautics and Space Administration, the Japanese Monbukagakusho,
 the Max Planck Society, and the Higher Education Funding Council for
 England. The SDSS Web Site is http://www.sdss.org/. The SDSS is
 managed by the Astrophysical Research Consortium for the
 Participating Institutions. The Participating Institutions are the
 American Museum of Natural History, Astrophysical Institute Potsdam,
 University of Basel, University of Cambridge, Case Western Reserve
 University, University of Chicago, Drexel University, Fermilab, the
 Institute for Advanced Study, the Japan Participation Group, Johns
 Hopkins University, the Joint Institute for Nuclear Astrophysics, the
 Kavli Institute for Particle Astrophysics and Cosmology, the Korean
 Scientist Group, the Chinese Academy of Sciences (LAMOST), Los Alamos
 National Laboratory, the Max-Planck-Institute for Astronomy (MPIA),
 the Max-Planck-Institute for Astrophysics (MPA), New Mexico State
 University, Ohio State University, University of Pittsburgh,
 University of Portsmouth, Princeton University, the United States
 Naval Observatory, and the University of Washington.

\bibliography{references}

%%%%%%%%%%%%%%%%%%%%%%
 \appendix 
\section{}

\subsection{Dependence on QSOs colour :}
Here, we present the dependece of \lya emission on background QSOs
colour by considering the $CNR \ge 4$ systems. The results are shown
in Fig.~\ref{fig:duststack_NHI21_cnr4_appx} and measured luminosities
are summarized in Table~\ref{tab:line_para_dust_withrelax_appdx} in
the Appendix. The double hump profile is clearly visible even in this
case that has $\sim$ 30 per cent more systems. However, the total
luminosity measured in the median stacked spectrum,
$(19.1\pm6.6)\times10^{40}$ erg s $^{-1}$, is slightly less but
consistent within 0.4$\sigma$ level to what we measure for $CNR \ge
5$. The double Gaussian fits are shown in
Fig.~\ref{fig:duststack_NHI21_cnr4_appx}. It is also clear from
Table~\ref{tab:line_para_dust_withrelax_appdx} that derived
luminosities using Gaussian fits to the blue and red hump are
significant at more than 3.3 $\sigma$ level. Here also we find
L$^r_\lambda$ being higher than L$^b_\lambda$. However, the excess is
less than what we have seen in the $CNR \ge 5$ sample. The measured
velocity shifts of Gaussian peaks with respect to the central
wavelength systemic redshift are consistent with that measured for the
$CNR \ge 5$ case. \emph{ This enlarged sample also shows : (i) the
  presence of enhanced emission in the blue part; (ii) the blue peak
  having more velocity shift with respect to the systemic redshift
  compared to the red and (iii) L$^r_\lambda$ being higher than
  L$^b_\lambda$ for systems with ($r-i$) $< 0.05$.} \par

  In addition, the trend of total luminosity being nearly same, and
  the double hump present only in low ($r-i$) sub-samples, as seen in
  Fig.~\ref{fig:duststack}, is also present when we consider systems
  with $CNR \ge 4$. However, the average luminosities are slightly
  less from what we find for $CNR \ge 5$ sample.

\begin{figure*}
 \epsfig{figure=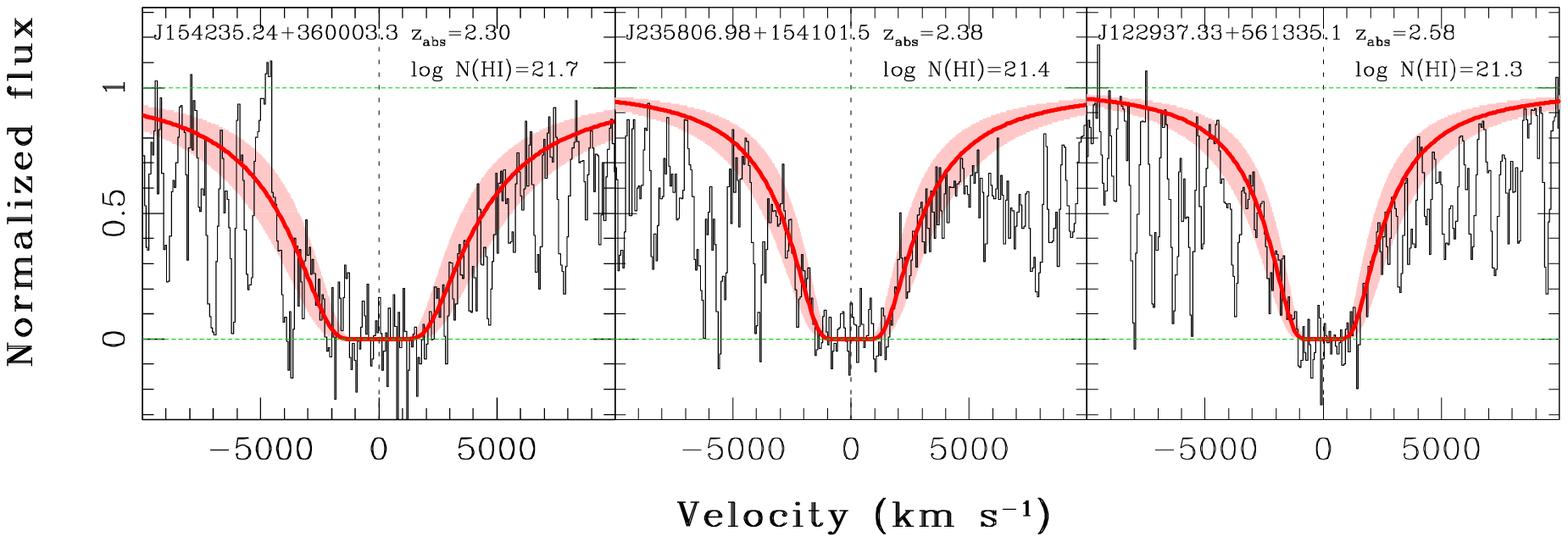,bbllx=32bp,bblly=163bp,bburx=564bp,bbury=351bp,clip=true,width=16.0cm,height=6.5cm}
 \caption{Examples of DLAs from our parent sample with large
   pixel-to-pixel variation in the absorption trough (see also,
   Section~\ref{subsec:compositespec}) }
 \label{fig:bad_pix_appx}
   \end{figure*}

\begin{figure*}
 \epsfig{figure=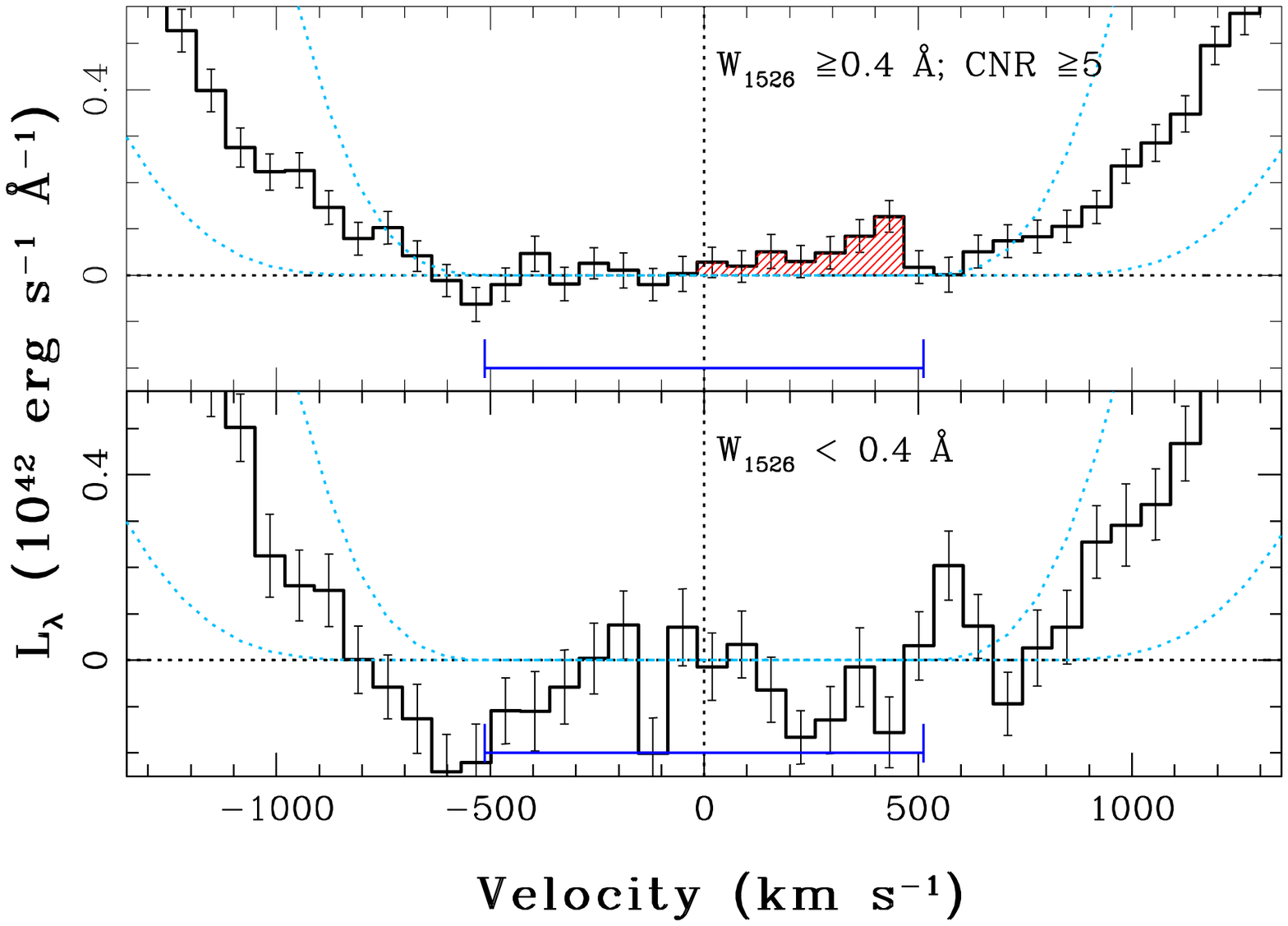,bbllx=22bp,bblly=144bp,bburx=568bp,bbury=535bp,clip=true,width=6.0cm,height=6.0cm}
 \epsfig{figure=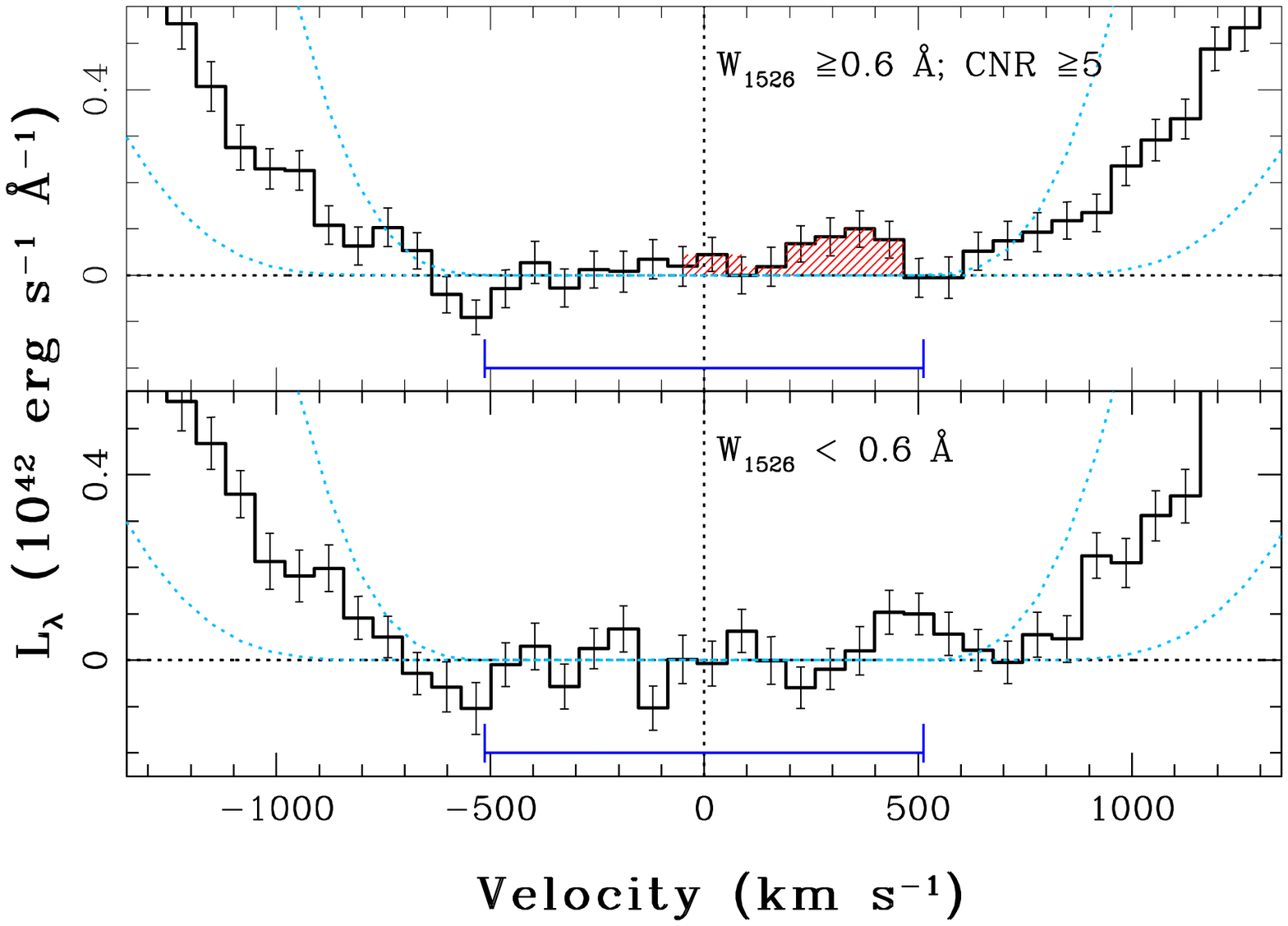,bbllx=60bp,bblly=144bp,bburx=568bp,bbury=535bp,clip=true,width=5.7cm,height=6.0cm}
 \epsfig{figure=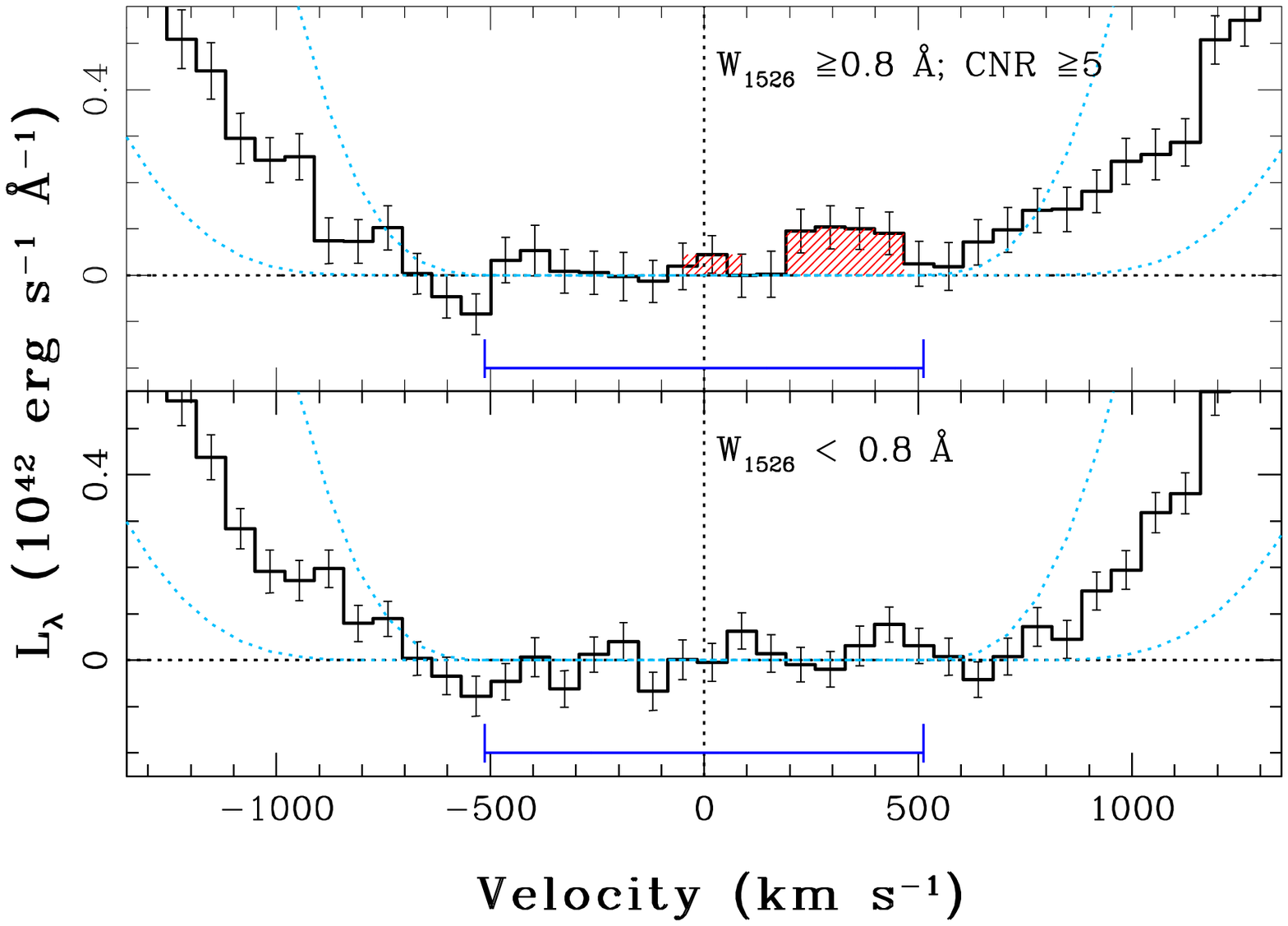,bbllx=60bp,bblly=144bp,bburx=568bp,bbury=535bp,clip=true,width=5.7cm,height=6.0cm}
 \caption{\emph{Left to right :} Comparison of median stacked spectrum
   for a sub-sample with CNR $\ge 5$ and $W_{1526}$ greater than
   (\emph{top panel}) and less than (\emph{bottom panels}) 0.4, 0.6
   and 0.8~\AA, respectively. The blue segment shows the DLA core with
   $\tau \ge 10$ for log~\nhi\ $\rm = 21.0$. The dashed curves
   show the synthetic profiles for lower (i.e., log~\nhi $ = \rm
   21)$ and median (i.e., log~\nhi $\rm = 21.23)$
   column density of DLAs used to get the stacked spectrum. }
 \label{fig:profile_cnr5_ewcomp_appx}
   \end{figure*}

\begin{figure*}
 \epsfig{figure=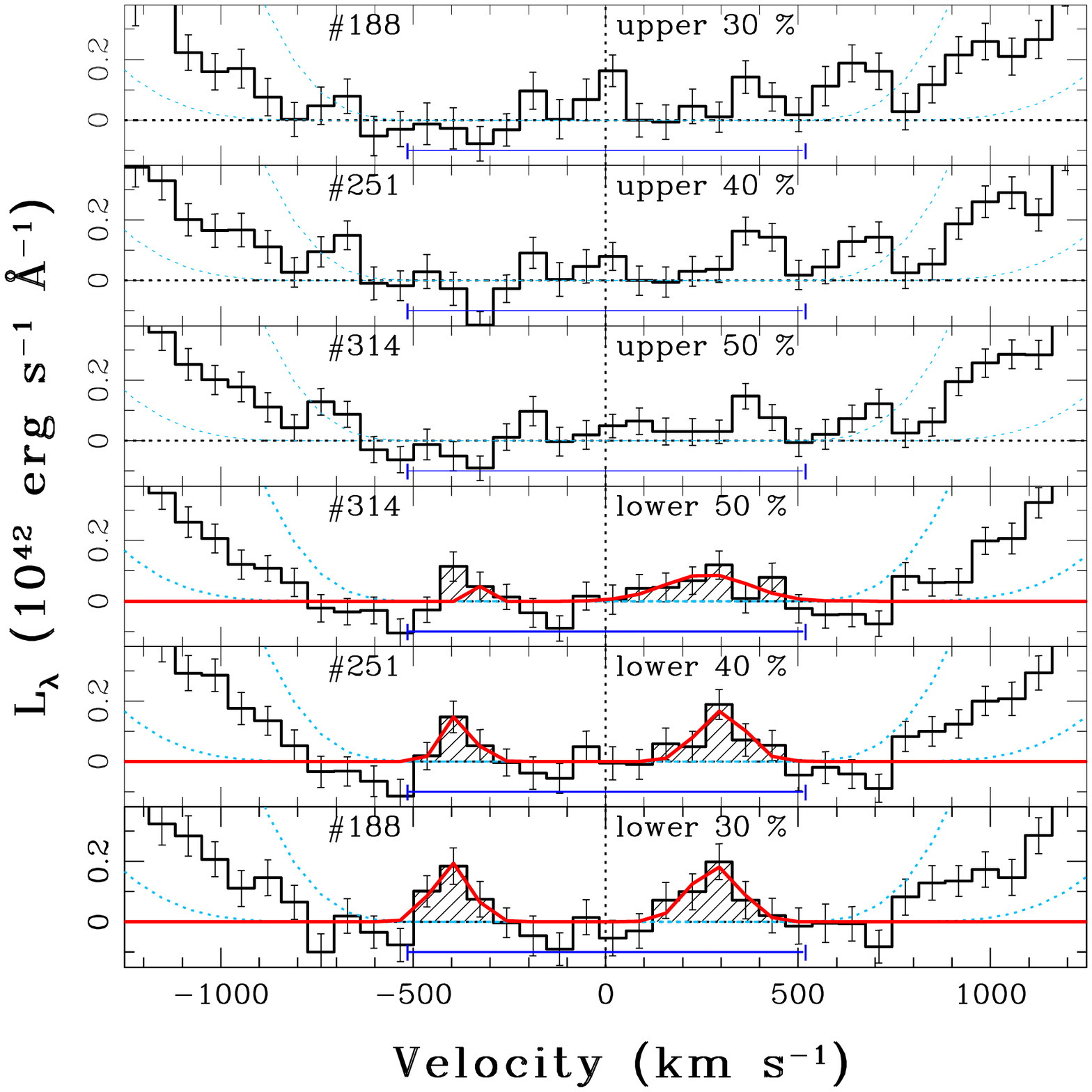,width=8.5cm,height=10.2cm}
 \epsfig{figure=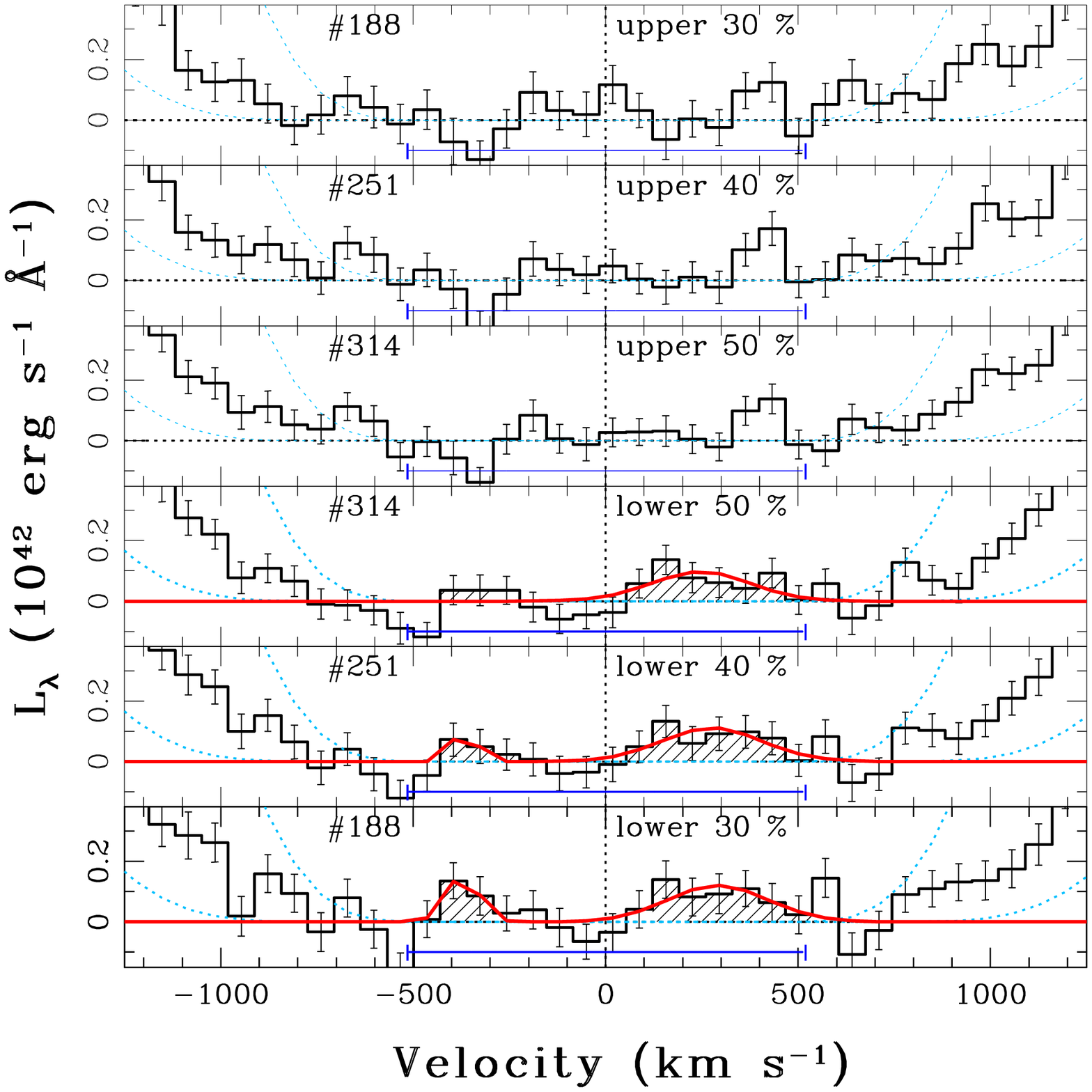,width=8.5cm,height=10.2cm}
 \caption{ Same as Fig.~\ref{fig:duststack}, for the
   sub-sample with relaxed CNR limit to CNR $\ge 4$.}
 \label{fig:duststack_NHI21_cnr4_appx}
   \end{figure*}

\begin{figure*}
 \epsfig{figure=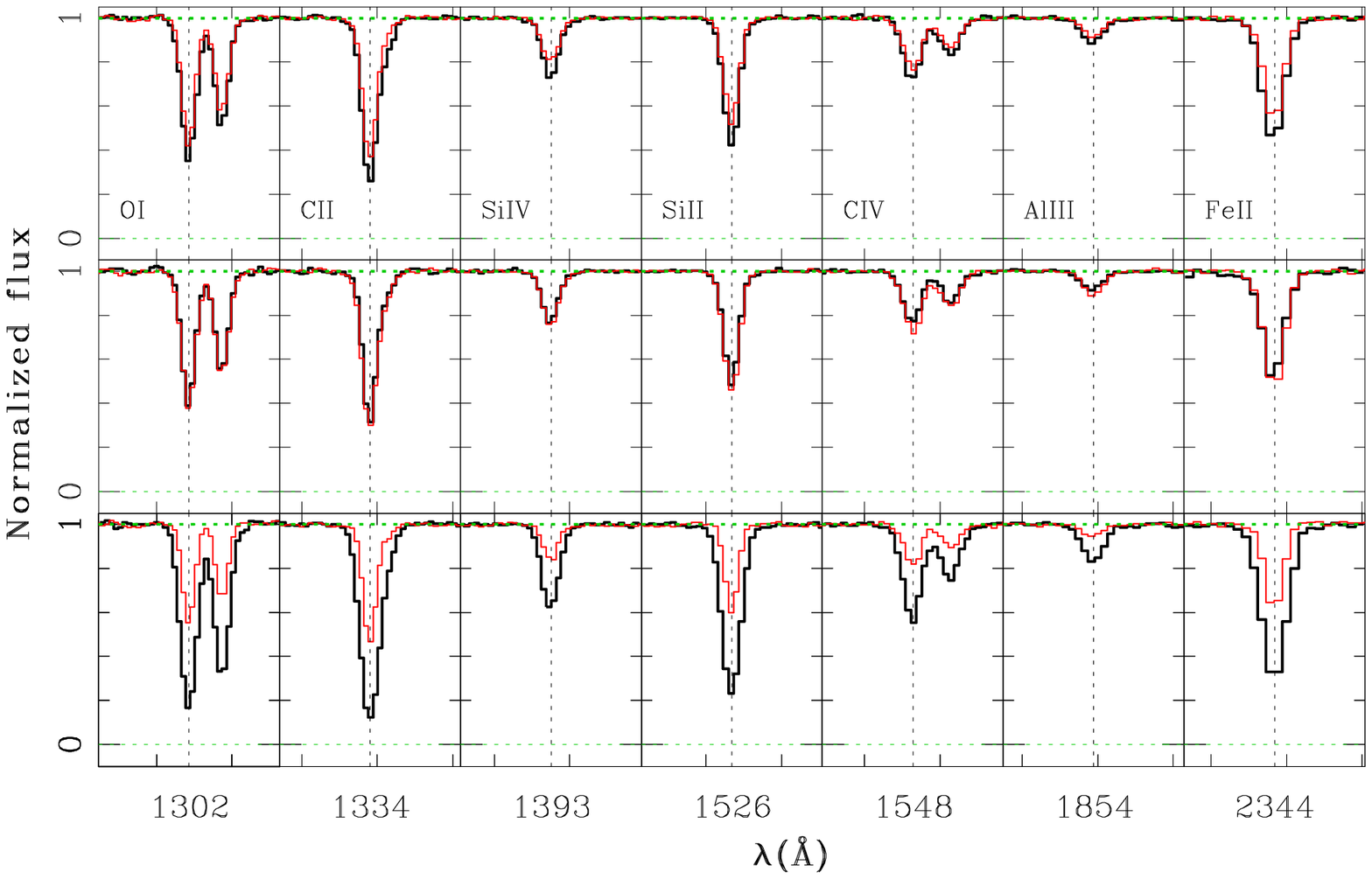,bbllx=39bp,bblly=162bp,bburx=567bp,bbury=500bp,clip=true,width=8.5cm,height=6.5cm}
 \epsfig{figure=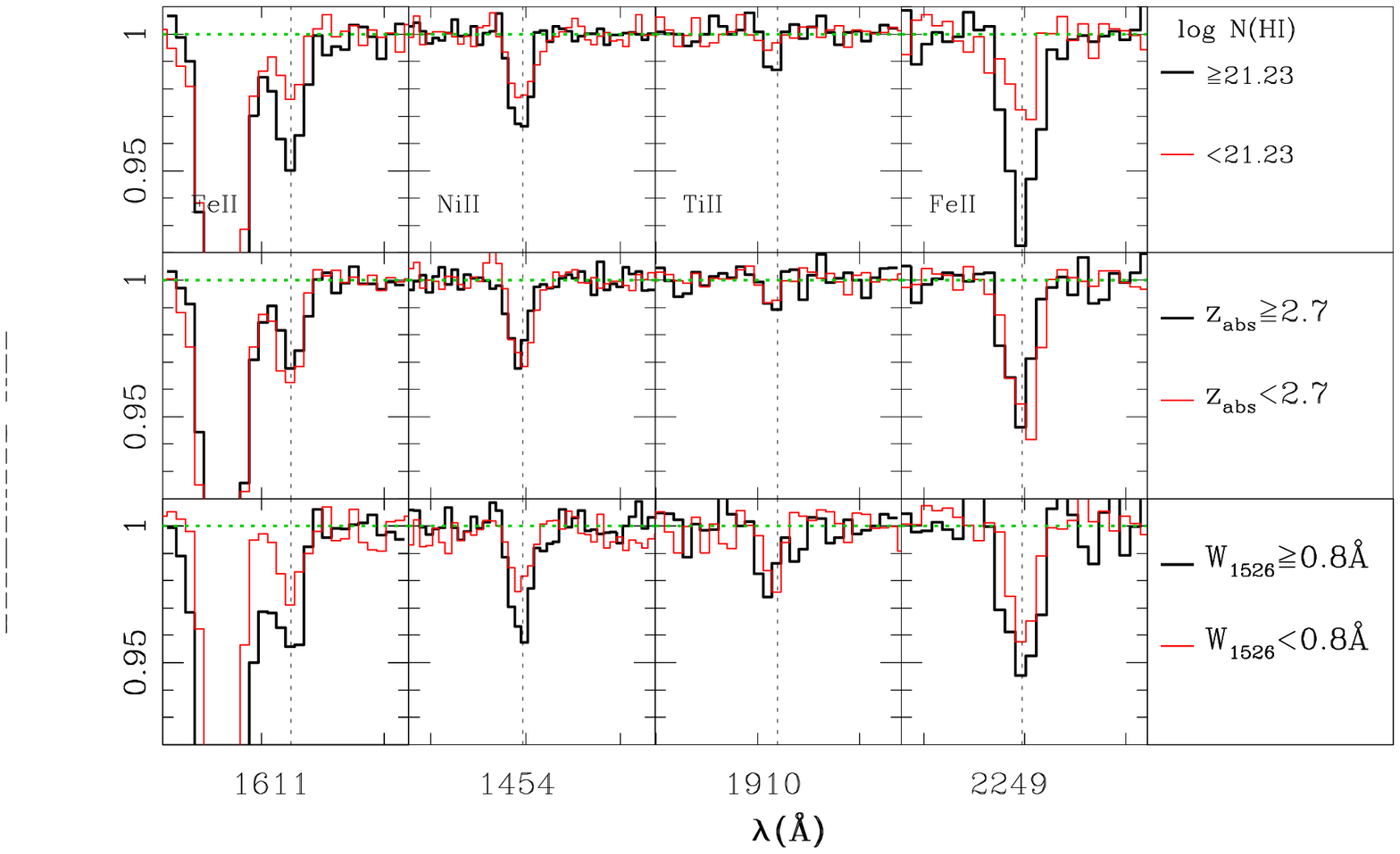,bbllx=62bp,bblly=162bp,bburx=567bp,bbury=500bp,clip=true,width=8.0cm,height=6.5cm}
 \caption{Comparison of several absorption lines in the composite
   spectra of a number of sub-samples, defined in
   Table~\ref{tab:line_lum_para}. Each row shows different lines in
   the composite spectra of the same two sub-samples (one in red and
   the other in black), which are detailed in the last column of each
   row. Each column contains lines of the same species, indicated in
   the top row of that column, the wavelength of which is indicated
   below each column.}
 \label{fig:vel_plot_appx}
   \end{figure*}

\begin{table*}
 \centering
 \begin{minipage}{160mm}
 {\small
 \caption{ \small{Luminosity of \lya line in bootstrap analysis of
     median stacked spectra for various sub-samples based on
     $W_{1526}$ and $CNR$ limit of $\ge 5$ and $\ge 4$.}}
 \label{tab:line_para_withrelax_apdx}
 \begin{tabular}{@{} l r r r r  r r r r c c@{}}
 \hline 
 \multicolumn{1}{c}{sample}  & \multicolumn{1}{c}{number} &\multicolumn{3}{c}{\lya luminosity ($\times 10^{40} \rm erg~s^{-1}$)}    \\
  \multicolumn{1}{c}{criteria}& &\multicolumn{3}{c}{median stack}\\
 
                          &   & DLA-bottom  & \multicolumn{1}{c}{L$_\lambda^b$}   &   \multicolumn{1}{c}{L$_\lambda^r$}        \\
      
\hline
$W_{1526} \ge 0.4 \rm \AA$  and $CNR \ge 5$  &513 & $ 13.49\pm  4.05$ & $  1.16\pm  2.84$ &$   12.34\pm  2.90$  \\
$W_{1526} <   0.4 \rm \AA$  and $CNR \ge 5$  &112 & $-26.27\pm  8.33$ & $-10.78\pm  5.92$ &$  -15.45\pm  5.85$  \\
$W_{1526} \ge 0.6 \rm \AA$  and $CNR \ge 5$  &391 & $ 13.76\pm  4.69$ & $  1.70\pm  3.00$ &$   12.05\pm  3.33$  \\
$W_{1526} <   0.6 \rm \AA$  and $CNR \ge 5$  &279 & $  3.41\pm  5.44$ & $ -1.43\pm  3.76$ &$    4.85\pm  3.93$  \\ \hline
$W_{1526} \ge 0.4 \rm \AA$  and $CNR \ge 4${\textcolor{blue}{$^a$}} & 627 & $ 11.05\pm  3.67$ & $  2.02\pm  1.65$ &$   11.48\pm  1.75$       \\
$W_{1526} <   0.4 \rm \AA$  and $CNR \ge 4$  &119 & $-28.18\pm  8.09$ & $-11.60\pm  5.78$ &$  -16.59\pm  5.66$  \\
$W_{1526} \ge 0.6 \rm \AA$  and $CNR \ge 4$  &496 & $  9.92\pm  4.18$ & $ -0.43\pm  2.93$ &$   10.34\pm  2.98$  \\
$W_{1526} <   0.6 \rm \AA$  and $CNR \ge 4$  &320 & $  1.87\pm  5.00$ & $ -2.90\pm  3.47$ &$    4.77\pm  3.60$  \\
$W_{1526} \ge 0.8 \rm \AA$  and $CNR \ge 4$  &370 & $  8.72\pm  4.78$ & $ -0.80\pm  3.36$ &$    9.52\pm  3.40$  \\
$W_{1526} <   0.8 \rm \AA$  and $CNR \ge 4$  &511 & $ -0.04\pm  4.03$ & $ -4.07\pm  2.80$ &$    4.03\pm  2.91$  \\
\hline                                                   
 \end{tabular} 
 }\\ 
 \end{minipage}
 \end{table*}

\begin{table*}
 \centering
\begin{minipage}{150mm}
 {\scriptsize
 \caption{ \small{Same as Table~\ref{tab:line_para_dust}, after
     relaxing the $CNR$ limit to $CNR \ge 4$.}}
 \label{tab:line_para_dust_withrelax_appdx}
 \begin{tabular}{@{} r r c c r r r   c c c @{}}
 \hline 
  \multicolumn{1}{c}{Sample}  & \multicolumn{1}{c}{($r-i$)}  & \multicolumn{1}{c}{$\langle \Delta (r-i) \rangle$}  & \multicolumn{3}{c}{\lya luminosity }  &\multicolumn{2}{c}{velocity-offset}  &FWHM\\
                             &&&  \multicolumn{3}{c}{($\times 10^{40} \rm erg~s^{-1}$)}&&&\multicolumn{1}{l}{$\rm ( km\ s^{-1})$}\\
 
                              &&& DLA-bottom  & \multicolumn{1}{c}{L$_\lambda^b$}   &   \multicolumn{1}{c}{L$_\lambda^r$} &  \multicolumn{1}{c}{$\Delta v^r$} & \multicolumn{1}{c}{$\Delta v^b$} &     \\
\hline
 
lower 30\% & $< 0.04$  &-0.08  & $ 19.07 \pm 6.61$&  $  7.65 \pm 4.42$& $ 11.41  \pm 4.92 $   & $283\pm  24$& $-401\pm 19$& $    -      $\\ 
lower 40\% & $< 0.07$  &-0.06  & $ 13.07 \pm 5.84$&  $  3.92 \pm 3.94$& $  9.15  \pm 4.31 $   & $301\pm  22$& $-383\pm 24$& $    -      $\\
lower 50\% & $< 0.10$  &-0.04  & $ 10.47 \pm 5.24$&  $  0.95 \pm 3.51$& $  9.52  \pm 3.89 $   & $263\pm  48$& $-336\pm 54$& $205 \pm112 $\\
upper 50\% & $> 0.10$  & 0.12  & $ 10.87 \pm 5.10$&  $ -0.65 \pm 3.66$& $ 11.52  \pm 3.56 $   & $     -    $& $      -   $& $      -    $\\
upper 40\% & $> 0.13$  & 0.13  & $ 13.27 \pm 5.64$&  $ -1.10 \pm 4.05$& $ 14.37  \pm 3.93 $   & $     -    $& $      -   $& $      -    $\\
upper 30\% & $> 0.16$  & 0.16  & $ 10.77 \pm 6.69$&  $ -0.38 \pm 4.93$& $ 11.15  \pm 4.52 $   & $     -    $& $      -   $& $      -    $\\

\hline                                                   
 \end{tabular} 
 } \\
 \end{minipage}
 \end{table*}

\begin{table*}
 \centering
 \begin{minipage}{180mm}
 {\scriptsize
 \caption{ \small {Equivalent widths of metal absorption lines for
     various sub-samples.}}
 \label{tab:line_EW}
 \begin{tabular}{@{} c c  c c c c c c  c@{}}
 \hline 
 \multicolumn{1}{c}{Transition} &\multicolumn{8}{c}{\ew (\AA)}     \\     
                                &\multicolumn{1}{c}{All }  & \multicolumn{3}{c}{W$_{1526}$}  &  \multicolumn{2}{c}{log\ \nhi}  & \multicolumn{2}{c}{ \zabs}   \\

                                &    & \multicolumn{1}{c}{$\ge 0.8$~\AA} & \multicolumn{1}{c}{$< 0.8$~\AA} & \multicolumn{1}{c}{$\ge 0.4$~\AA} & \multicolumn{1}{c}{$ \ge 21.23 $} & \multicolumn{1}{c}{$ < 21.23 $} & \multicolumn{1}{c}{$ \ge 2.7$}& \multicolumn{1}{c}{$ < 2.7$} \\                                
                                &(S1) & (S2) & (S3)     & (S4)   &  (S5) & (S6)  &  (S7) &  (S8)              \\
\hline

Al{\sc~ii}$~\lambda$1670 	     & 0.732$\pm$ 0.008&  1.218$\pm$ 0.012 & 0.483 $\pm$  0.008 &  0.864$\pm$  0.009 & 0.841$\pm$  0.012 &  0.589$\pm$  0.011 & 0.713$\pm$  0.011 & 0.735$\pm$  0.012 \\
Al{\sc~iii}$~\lambda$1854	     & 0.171$\pm$ 0.006&  0.332$\pm$ 0.009 & 0.092 $\pm$  0.007 &  0.219$\pm$  0.007 & 0.200$\pm$  0.008 &  0.153$\pm$  0.008 & 0.161$\pm$  0.009 & 0.175$\pm$  0.008 \\
Al{\sc~iii}$~\lambda$1862	     & 0.103$\pm$ 0.006&  0.192$\pm$ 0.009 & 0.050 $\pm$  0.007 &  0.135$\pm$  0.006 & 0.134$\pm$  0.008 &  0.079$\pm$  0.008 & 0.084$\pm$  0.009 & 0.105$\pm$  0.007 \\ \hline
                          	            	            		     	                         	             		         	              		      	  		       
Co{\sc~ii}$~\lambda$1466 	     & 0.009$\pm$ 0.003&  0.022$\pm$ 0.004 & 0.002 $\pm$  0.004 &  0.012$\pm$  0.003 & 0.007$\pm$  0.004 &  0.002$\pm$  0.004 & 0.011$\pm$  0.005 & 0.008$\pm$  0.003 \\ \hline
                          	            	            		     	     	                 	             		         	              		      	  		       
Cr{\sc~ii}$~\lambda$2056 	     & 0.085$\pm$ 0.007&  0.142$\pm$ 0.010 & 0.060 $\pm$  0.009 &  0.077$\pm$  0.007 & 0.087$\pm$  0.009 &  0.038$\pm$  0.010 & 0.056$\pm$  0.012 & 0.093$\pm$  0.007 \\
Cr{\sc~ii}$~\lambda$2066 	     & 0.043$\pm$ 0.006&  0.089$\pm$ 0.009 & 0.048 $\pm$  0.008 &  0.047$\pm$  0.007 & 0.032$\pm$  0.008 &  0.039$\pm$  0.009 & 0.043$\pm$  0.010 & 0.054$\pm$  0.007 \\ \hline
                          	            	            		     	                         	             		         	              		      	  		       
C{\sc~ii}$~\lambda$1334  	     & 0.934$\pm$ 0.010&  1.513$\pm$ 0.015 & 0.644 $\pm$  0.011 &  1.167$\pm$  0.012 & 1.094$\pm$  0.015 &  0.812$\pm$  0.014 & 0.866$\pm$  0.015 & 0.977$\pm$  0.014 \\
C{\sc~iv}$~\lambda$1548  	     & 0.362$\pm$ 0.007&  0.676$\pm$ 0.012 & 0.253 $\pm$  0.007 &  0.508$\pm$  0.008 & 0.399$\pm$  0.009 &  0.367$\pm$  0.010 & 0.321$\pm$  0.010 & 0.420$\pm$  0.009 \\
C{\sc~iv}$~\lambda$1550  	     & 0.228$\pm$ 0.006&  0.403$\pm$ 0.011 & 0.171 $\pm$  0.007 &  0.305$\pm$  0.007 & 0.247$\pm$  0.008 &  0.234$\pm$  0.009 & 0.175$\pm$  0.009 & 0.264$\pm$  0.008 \\ \hline
                          	            	            		     	                         	             		         	              		      	  		       
Fe{\sc~ii}$~\lambda$1608 	     & 0.456$\pm$ 0.006&  0.750$\pm$ 0.010 & 0.293 $\pm$  0.007 &  0.569$\pm$  0.007 & 0.534$\pm$  0.009 &  0.411$\pm$  0.008 & 0.432$\pm$  0.009 & 0.494$\pm$  0.009 \\
Fe{\sc~ii}$~\lambda$1611 	     & 0.048$\pm$ 0.004&  0.070$\pm$ 0.005 & 0.028 $\pm$  0.005 &  0.056$\pm$  0.004 & 0.078$\pm$  0.005 &  0.025$\pm$  0.005 & 0.040$\pm$  0.005 & 0.050$\pm$  0.005 \\
Fe{\sc~ii}$~\lambda$2249 	     & 0.072$\pm$ 0.010&  0.164$\pm$ 0.015 & 0.062 $\pm$  0.014 &  0.105$\pm$  0.011 & 0.158$\pm$  0.014 &  0.053$\pm$  0.015 & 0.094$\pm$  0.018 & 0.102$\pm$  0.011 \\
Fe{\sc~ii}$~\lambda$2260 	     & 0.089$\pm$ 0.009&  0.156$\pm$ 0.014 & 0.078 $\pm$  0.012 &  0.110$\pm$  0.010 & 0.132$\pm$  0.013 &  0.080$\pm$  0.013 & 0.076$\pm$  0.017 & 0.124$\pm$  0.010 \\
Fe{\sc~ii}$~\lambda$2344 	     & 0.894$\pm$ 0.023&  1.471$\pm$ 0.029 & 0.578 $\pm$  0.030 &  1.152$\pm$  0.022 & 1.083$\pm$  0.031 &  0.819$\pm$  0.030 & 0.886$\pm$  0.045 & 0.891$\pm$  0.019 \\
Fe{\sc~ii}$~\lambda$2374 	     & 0.605$\pm$ 0.020&  0.988$\pm$ 0.028 & 0.427 $\pm$  0.026 &  0.704$\pm$  0.022 & 0.744$\pm$  0.029 &  0.477$\pm$  0.027 & 0.565$\pm$  0.041 & 0.622$\pm$  0.018 \\
Fe{\sc~ii}$~\lambda$2382 	     & 1.221$\pm$ 0.021&  1.902$\pm$ 0.030 & 0.894 $\pm$  0.026 &  1.384$\pm$  0.024 & 1.329$\pm$  0.030 &  1.075$\pm$  0.030 & 1.142$\pm$  0.043 & 1.223$\pm$  0.020 \\
Fe{\sc~ii}$~\lambda$2586 	     & 0.791$\pm$ 0.041&  1.363$\pm$ 0.061 & 0.532 $\pm$  0.047 &  1.038$\pm$  0.043 & 0.913$\pm$  0.053 &  0.706$\pm$  0.058 & 0.587$\pm$  0.120 & 0.856$\pm$  0.030 \\
Fe{\sc~ii}$~\lambda$2600 	     & 1.186$\pm$ 0.042&  1.895$\pm$ 0.059 & 0.778 $\pm$  0.051 &  1.448$\pm$  0.042 & 1.353$\pm$  0.053 &  1.069$\pm$  0.062 & 0.998$\pm$  0.127 & 1.265$\pm$  0.032 \\ \hline
                          	            	            		     	                         	             		         	              		      	  		       
Mg{\sc~ii}$~\lambda$2796 	     & 1.901$\pm$ 0.086&  2.954$\pm$ 0.111 & 1.267 $\pm$  0.116 &  2.184$\pm$  0.084 & 2.188$\pm$  0.098 &  1.689$\pm$  0.134 & $-$  $\pm$  $-$   & 1.904$\pm$  0.084 \\
Mg{\sc~ii}$~\lambda$2803 	     & 1.813$\pm$ 0.105&  2.858$\pm$ 0.140 & 1.123 $\pm$  0.104 &  2.107$\pm$  0.094 & 2.010$\pm$  0.097 &  1.716$\pm$  0.150 & $-$  $\pm$  $-$   & 1.828$\pm$  0.105 \\
Mg{\sc~i}$~\lambda$2852  	     & 0.418$\pm$ 0.075&  0.712$\pm$ 0.100 & 0.254 $\pm$  0.112 &  0.586$\pm$  0.083 & 0.560$\pm$  0.103 &  0.326$\pm$  0.105 & $-$  $\pm$  $-$   & 0.416$\pm$  0.076 \\ \hline
                          	            	            		     	                         	             		         	              		      	  		       
Mn{\sc~ii}$~\lambda$2576 	     & 0.142$\pm$ 0.033&  0.169$\pm$ 0.054 & 0.121 $\pm$  0.032 &  0.157$\pm$  0.037 & 0.176$\pm$  0.029 &  0.107$\pm$  0.062 & 0.064$\pm$  0.103 & 0.155$\pm$  0.018 \\
Mn{\sc~ii}$~\lambda$2594 	     & 0.057$\pm$ 0.031&  0.105$\pm$ 0.052 & 0.049 $\pm$  0.034 &  0.115$\pm$  0.035 & 0.108$\pm$  0.039 &  0.037$\pm$  0.042 & 0.010$\pm$  0.103 & 0.093$\pm$  0.018 \\
Mn{\sc~ii}$~\lambda$2606 	     & 0.032$\pm$ 0.026&  0.083$\pm$ 0.039 & 0.010 $\pm$  0.033 &  0.061$\pm$  0.028 & 0.096$\pm$  0.036 &  0.010$\pm$  0.036 & 0.021$\pm$  0.084 & 0.055$\pm$  0.018 \\ \hline
                          	            	            		     	                         	             		         	              		      	  		       
Ni{\sc~ii}$~\lambda$1317 	     & 0.131$\pm$ 0.009&  0.209$\pm$ 0.015 & 0.096 $\pm$  0.011 &  0.121$\pm$  0.011 & 0.123$\pm$  0.012 &  0.111$\pm$  0.013 & 0.109$\pm$  0.013 & 0.157$\pm$  0.012 \\
Ni{\sc~ii}$~\lambda$1370 	     & 0.057$\pm$ 0.005&  0.078$\pm$ 0.008 & 0.034 $\pm$  0.007 &  0.083$\pm$  0.006 & 0.105$\pm$  0.007 &  0.012$\pm$  0.007 & 0.056$\pm$  0.007 & 0.078$\pm$  0.008 \\
Ni{\sc~ii}$~\lambda$1454 	     & 0.029$\pm$ 0.003&  0.049$\pm$ 0.005 & 0.027 $\pm$  0.004 &  0.044$\pm$  0.004 & 0.039$\pm$  0.005 &  0.026$\pm$  0.005 & 0.034$\pm$  0.005 & 0.031$\pm$  0.004 \\
Ni{\sc~ii}$~\lambda$1467 	     & 0.024$\pm$ 0.003&  0.055$\pm$ 0.005 & 0.009 $\pm$  0.004 &  0.039$\pm$  0.004 & 0.025$\pm$  0.005 &  0.014$\pm$  0.005 & 0.031$\pm$  0.005 & 0.026$\pm$  0.004 \\
Ni{\sc~ii}$~\lambda$1709 	     & 0.059$\pm$ 0.004&  0.095$\pm$ 0.007 & 0.036 $\pm$  0.006 &  0.083$\pm$  0.005 & 0.087$\pm$  0.006 &  0.039$\pm$  0.006 & 0.038$\pm$  0.006 & 0.073$\pm$  0.006 \\
Ni{\sc~ii}$~\lambda$1741 	     & 0.079$\pm$ 0.005&  0.144$\pm$ 0.008 & 0.054 $\pm$  0.007 &  0.101$\pm$  0.006 & 0.111$\pm$  0.007 &  0.035$\pm$  0.007 & 0.070$\pm$  0.007 & 0.078$\pm$  0.007 \\
Ni{\sc~ii}$~\lambda$1751 	     & 0.063$\pm$ 0.005&  0.069$\pm$ 0.007 & 0.031 $\pm$  0.006 &  0.067$\pm$  0.005 & 0.072$\pm$  0.006 &  0.023$\pm$  0.007 & 0.059$\pm$  0.006 & 0.034$\pm$  0.007 \\ \hline
                          	            	            		     	                         	             		         	              		      	  		       
O{\sc~i}$~\lambda$1302   	     & 0.732$\pm$ 0.009&  1.186$\pm$ 0.013 & 0.457 $\pm$  0.011 &  0.903$\pm$  0.010 & 0.794$\pm$  0.013 &  0.669$\pm$  0.013 & 0.715$\pm$  0.013 & 0.724$\pm$  0.013 \\ \hline
                          	            	            		     	                         	             		         	              		      	  		       
Si{\sc~ii}$~\lambda$1304 	     & 0.546$\pm$ 0.009&  0.930$\pm$ 0.014 & 0.341 $\pm$  0.012 &  0.649$\pm$  0.011 & 0.592$\pm$  0.014 &  0.520$\pm$  0.013 & 0.531$\pm$  0.014 & 0.540$\pm$  0.013 \\
Si{\sc~ii}$~\lambda$1526 	     & 0.674$\pm$ 0.007&  1.206$\pm$ 0.010 & 0.427 $\pm$  0.007 &  0.846$\pm$  0.008 & 0.752$\pm$  0.010 &  0.606$\pm$  0.010 & 0.649$\pm$  0.011 & 0.694$\pm$  0.010 \\
Si{\sc~ii}$~\lambda$1808 	     & 0.141$\pm$ 0.006&  0.262$\pm$ 0.009 & 0.090 $\pm$  0.007 &  0.207$\pm$  0.006 & 0.221$\pm$  0.008 &  0.100$\pm$  0.008 & 0.142$\pm$  0.008 & 0.159$\pm$  0.008 \\ \hline
                          	            	            		     	                         	             		         	              		      	  		       
Si{\sc~iv}$~\lambda$1393 	     & 0.321$\pm$ 0.006&  0.543$\pm$ 0.010 & 0.208 $\pm$  0.007 &  0.373$\pm$  0.007 & 0.342$\pm$  0.009 &  0.268$\pm$  0.008 & 0.312$\pm$  0.009 & 0.331$\pm$  0.008 \\
Si{\sc~iv}$~\lambda$1402 	     & 0.169$\pm$ 0.005&  0.356$\pm$ 0.008 & 0.120 $\pm$  0.006 &  0.231$\pm$  0.006 & 0.190$\pm$  0.007 &  0.158$\pm$  0.007 & 0.183$\pm$  0.007 & 0.204$\pm$  0.007 \\ \hline
                          	            	            		     	                         	             		         	              		      	  		       
Ti{\sc~ii}$~\lambda$1910 	     & 0.021$\pm$ 0.005&  0.039$\pm$ 0.007 & 0.015 $\pm$  0.006 &  0.037$\pm$  0.005 & 0.030$\pm$  0.007 &  0.011$\pm$  0.007 & 0.015$\pm$  0.008 & 0.017$\pm$  0.006 \\ \hline
				            	            		     	                         	             		         	              		      	  		       
Zn{\sc~ii}+Mg{\sc~i}$~\lambda$2026   & 0.111$\pm$ 0.007&  0.179$\pm$ 0.010 & 0.057 $\pm$  0.008 &  0.120$\pm$  0.008 & 0.137$\pm$  0.010 &  0.070$\pm$  0.009 & 0.082$\pm$  0.011 & 0.097$\pm$  0.008 \\
Zn{\sc~ii}+Cr{\sc~ii}$~\lambda$2062  & 0.102$\pm$ 0.006&  0.189$\pm$ 0.010 & 0.079 $\pm$  0.008 &  0.123$\pm$  0.007 & 0.116$\pm$  0.009 &  0.061$\pm$  0.009 & 0.079$\pm$  0.011 & 0.129$\pm$  0.007 \\ \hline
 \hline   
   
 \end{tabular} 
 }   
 \end{minipage}
 \end{table*}

\label{lastpage}

\end{document}